\documentclass[a4paper,12pt]{article}
\pdfoutput=1

\usepackage{jcappub}
\usepackage[symbol]{footmisc}
\usepackage{amsmath,amssymb}
\usepackage{graphicx}
\usepackage{mathrsfs}
\usepackage{adjustbox}
\usepackage{multirow}
\usepackage[normalem]{ulem}
\usepackage{bm}
\usepackage{orcidlink}

\usepackage{tabulary}
\usepackage{color}

\newcommand{\nc}{\newcommand*}

\nc{\al}{\alpha}
\nc{\s}{\sigma}
\nc{\dt}{\delta}
\nc{\Dt}{\Delta}
\nc{\Ld}{\Lambda}
\nc{\p}{\partial}
\nc{\om}{\omega}
\nc{\Om}{\Omega}
\nc{\rd}{\mathrm{d}}
\nc{\Od}[1]{\mathcal{O}(#1)}
\nc{\kp}{\kappa}

\def\e{\begin{equation}}
\def\q{\end{equation}}
\def\m{\begin{eqnarray}}
\def\n{\end{eqnarray}}

\nc{\Eq}[1]{Eq.~\eqref{#1}}
\nc{\Fig}[1]{Fig.~\ref{#1}}
\nc{\Table}[1]{Table~\ref{#1}}
\nc{\Sec}[1]{Sec.~\ref{#1}}

\newcommand{\PR}{{\cal P}_{\cal R}}
\newcommand{\OGW}{\Omega_{\rm GW}}
\newcommand{\dd}{\mathrm{d}}
\newcommand{\AR}{{\cal A}_{\cal R}}
\newcommand{\Heavi}{\Theta}
\newcommand{\Em}{{\cal E}_-}
\newcommand{\Ep}{{\cal E}_+}

\newcommand{\OGWx}[1]{\Omega_{{\rm GW},#1}}

\newcommand{\Csi}{C_{\rm SI}}
\nc{\red}[1]{\textcolor{red}{#1}}

\begin{document}

\title{Scalar-induced gravitational waves from a box-shaped curvature power spectrum}

\author[a,b]{Zu-Cheng Chen\orcidlink{0000-0001-7016-9934}}
\author[c,*]{Lang~Liu,\note{Corresponding author.}\orcidlink{0000-0002-0297-9633}}

\affiliation[a]{Department of Physics and Synergetic Innovation Center for Quantum Effects and Applications, Hunan Normal University, Changsha, Hunan 410081, China}
\affiliation[b]{Institute of Interdisciplinary Studies, Hunan Normal University, Changsha, Hunan 410081, China}
\affiliation[c]{Faculty of Arts and Sciences, Beijing Normal University, Zhuhai 519087, China}

\emailAdd{zuchengchen@gmail.com}
\emailAdd{liulang@bnu.edu.cn}

\abstract{
We compute the stochastic background of gravitational waves (GWs) produced at
second order in cosmological perturbation theory by a primordial curvature
power spectrum that is flat in $\ln k$ over a finite band
$[k_-,k_+]=[k_*e^{-\Delta},k_*e^{+\Delta}]$ and vanishes elsewhere.  This
logarithmic box interpolates between a monochromatic spectrum as
$\Delta\to0$ and a locally scale-invariant plateau as $\Delta\to\infty$, and
its sharp boundaries make the geometry of the source convolution unusually
transparent.  Working in the radiation era, we reduce the convolution to a
compact integral over the overlap between the momentum triangle and the box,
and evaluate it analytically in two regimes.  For a narrow box we show that,
to leading order in the width, the  spectrum equals the Dirac-spectrum result
multiplied by a purely geometric overlap factor $\Phi_{\rm box}(\kappa,\Delta)$;
this factor turns the infrared slope from $k^{3}\ln^{2}k$ into $k^{2}\ln^{2}k$
at a break $k_b=2k_*\sinh\Delta$.  For a broad box we separate the lower edge,
the scale-invariant interior, and the upper edge, derive the leading behaviour
in each (an infrared $k^{3}\ln^{2}k$ rise, a scale-invariant plateau, and a quartic cut-off at the hard endpoint $k=2k_+$), and
combine them into a single uniform formula cast as a product of two
universal, $\Delta$-independent edge functions.  We also provide an
integral-free closed-form surrogate for these edge functions for use in
parameter scans.  %Comparing with the lognormal peak, the geometric factor $\Phi_{\rm box}$ plays the role of the error function, the break-to-peak frequency ratio $f_b/f_p\simeq\sqrt3\,\Delta$ is shared by both shapes, and a genuine difference appears only in the broad limit, where the box develops a flat plateau and a hard ultraviolet edge rather than smooth tails.
}

\maketitle
%%%%%%%%%%%%%%%%%%%%%%%%%%%%%%%%%%%%%%%%%%%%%%%%%%%%%%%%%%%%%%%%%

\section{Introduction}

The observation of gravitational-wave (GW) transients from compact-binary
coalescences has turned GWs into a working observational tool, and the search
for stochastic GW backgrounds is now a central target of present and
planned detectors. Among the cosmological backgrounds, the one produced at
second order in cosmological perturbation theory, the GWs induced by linear
scalar perturbations, stands out because its source is the same
small-scale curvature perturbation whose amplitude is otherwise almost
unconstrained below the megaparsec scale. At linear order the tensor and scalar
sectors evolve independently in a Friedmann background, but the quadratic terms
of the Einstein equations couple them, so a sufficiently large scalar power
sources a tensor background whose amplitude and shape mirror those of the scalar
spectrum~\cite{Tomita:1967wkp,Matarrese:1992rp,Matarrese:1993zf,Matarrese:1997ay,Mollerach:2003nq,Carbone:2004iv,Ananda:2006af,Baumann:2007zm,Saito:2008jc,Kohri:2018awv,Espinosa:2018eve,Domenech:2021ztg}.
Detailed studies of the induced spectrum, its amplitude, its infrared scaling
and its sensitivity to the expansion history, have since been developed
extensively~\cite{Assadullahi:2009nf,Bugaev:2009zh,Bugaev:2010bb,Alabidi:2012ex,Alabidi:2013lya,Nakama:2016gzw,Inomata:2018epa,Yuan:2019udt,Piccoli:2026grg}.

The same enhanced small-scale curvature power is the prerequisite for the
gravitational collapse of rare overdensities into primordial black holes
(PBHs)~\cite{Hawking:1971ei,Carr:1974nx,Carr:1975qj,Ivanov:1994pa,Garcia-Bellido:1996mdl,Carr:2009jm,Carr:2016drx,Carr:2020gox}
at horizon re-entry. Since both phenomena are seeded by the curvature
perturbation at essentially the same comoving scale, the induced GW background
acts as an almost inevitable companion of any PBH-forming scenario, and the two
observables together sharpen the reconstruction of the primordial spectrum on
small scales~\cite{Saito:2008jc,Cai:2018dig,Sasaki:2018dmp,LISACosmologyWorkingGroup:2025vdz}. PBHs are, in
addition, a natural dark-matter candidate and may account for a fraction of the
binary black holes observed by ground-based
interferometers~\cite{Sasaki:2016jop,Bird:2016dcv,Clesse:2016vqa,Carr:2016drx,Ali-Haimoud:2017rtz,Raidal:2017mfl,Chen:2018czv,Liu:2018ess,Liu:2019rnx,Vaskonen:2019jpv,DeLuca:2020qqa,Hutsi:2020sol,Franciolini:2021tla};
see Refs.~\cite{Sasaki:2018dmp,Carr:2020gox,Green:2020jor,Escriva:2022duf,Carr:2020xqk,Yuan:2021qgz,Villanueva-Domingo:2021spv}
for reviews. The mapping from the curvature spectrum to the PBH abundance
depends sensitively on the collapse threshold and on the statistics of the
fluctuations~\cite{Young:2014ana,Germani:2018jgr,Musco:2018rwt,Escriva:2019phb,Carr:2009jm,Niikura:2017zjd,Musco:2020jjb}.
This dual role has motivated a large body of work mapping out the GW signatures
of inflationary mechanisms that boost the curvature spectrum at a chosen
wavenumber~\cite{Garcia-Bellido:2017mdw,Pi:2017gih,Inomata:2017okj,Espinosa:2017sgp,Kannike:2017bxn,Cai:2018tuh,Byrnes:2018txb,Motohashi:2017kbs,Ezquiaga:2017fvi,Di:2017ndc,Dalianis:2018frf,Bhaumik:2019tvl,Mishra:2019pzq,Liu:2020oqe,Fu:2019ttf,Fu:2020lob,Inomata:2021uqj,Kawai:2021edk,Ballesteros:2020qam,Ragavendra:2020sop,Yogesh:2025hll,Ozsoy:2018flq,Cicoli:2018asa,Ozsoy:2023ryl,Pattison:2017mbe,Biagetti:2018pjj,Ezquiaga:2019ftu,Cole:2017gle,Mohammadi:2025avz}.
Depending on the scale of the enhancement, the induced background falls within
the reach of pulsar timing arrays or of planned interferometers such as
LISA~\cite{LISA:2017pwj}, DECIGO~\cite{Kawamura:2011zz,Kawamura:2020pcg}, the
Einstein Telescope~\cite{Punturo:2010zz,Maggiore:2019uih}, Cosmic
Explorer~\cite{Reitze:2019iox}, BBO~\cite{Crowder:2005nr},
Taiji~\cite{Ruan:2018tsw}, TianQin~\cite{TianQin:2015yph} and the
SKA~\cite{Janssen:2014dka,Hobbs:2009yy}. This prospect has become especially topical since several pulsar-timing-array collaborations reported evidence for a nanohertz stochastic background~\cite{NANOGrav:2023gor,EPTA:2023fyk,Reardon:2023gzh,Xu:2023wog,Grunthal:2024sor}, for which the scalar-induced-GW mechanism provides a competitive interpretation~\cite{Chen:2019xse, NANOGrav:2023hvm,Dandoy:2023jot,Franciolini:2023pbf,Franciolini:2023wjm,Inomata:2023zup,Cai:2023dls,Wang:2023ost,Liu:2023ymk,Unal:2023srk,Figueroa:2023zhu,Zhu:2023faa,Firouzjahi:2023lzg,Li:2023qua,You:2023rmn,Balaji:2023ehk,HosseiniMansoori:2023mqh,Zhao:2023joc,Liu:2023pau,Yi:2023tdk,Bhaumik:2023wmw,Choudhury:2023hfm,Yi:2023npi,Harigaya:2023pmw,Jin:2023wri,Cannizzaro:2023mgc,Zhang:2023nrs,Liu:2023hpw,Domenech:2024rks,Choudhury:2023fwk,Tagliazucchi:2023dai,Basilakos:2023jvp,Inomata:2023drn,Li:2023xtl,Domenech:2023dxx,Gangopadhyay:2023qjr,Cyr:2023pgw,Chen:2024fir,Chen:2024twp,Choudhury:2023fjs,Choudhury:2024one,Cai:2025ksu}.

Because the induced spectrum is a quadratic functional of the curvature
spectrum, its shape is a fingerprint of the source. Reusable analytic templates
are therefore valuable for two reasons: they make the physics transparent, and
they remove the cost of repeated numerical convolutions in the parameter scans
required for forecasting and inference. Closed or semi-closed expressions are by
now available for a monochromatic (delta) source~\cite{Kohri:2018awv}, for a
lognormal peak~\cite{Pi:2020otn}, and for a broken power-law
peak~\cite{Li:2024lxx}. A recurring lesson of these
studies is that the infrared (IR) tail is governed not by the detail of the peak
but by its causal, support-related properties: a strictly monochromatic source
gives a universal $k^2$ slope, whereas any finite width turns the far IR into
$k^3$, with a logarithmic running on top~\cite{Cai:2019cdl,Yuan:2019wwo}.
Throughout we use the oscillation-averaged, sub-horizon spectrum in the
radiation era, whose gauge dependence has been examined in detail in
Refs.~\cite{Inomata:2019yww,DeLuca:2019ufz,Yuan:2019fwv,Hwang:2017oxa,Tomikawa:2019tvi,Domenech:2020xin,Yuan:2024qfz,Yuan:2025seu,Domenech:2025ccu},
and we take the curvature perturbation to be Gaussian, referring the reader to
Refs.~\cite{Cai:2018dig,Unal:2018yaa,Atal:2021jyo,Franciolini:2018vbk,Atal:2019cdz,Adshead:2021hnm,Iovino:2024sgs, Perna:2026szd, Iovino:2025cdy, Ferrante:2022mui,Ragavendra:2021qdu,Perna:2024ehx,Yuan:2023ofl,Garcia-Saenz:2022tzu}
for the impact of primordial non-Gaussianity.

In this paper we add to this catalogue the simplest source that carries 
sharp spectral edges: a power spectrum that is flat in $\ln k$ over a band of
half-width $\Delta$ and zero elsewhere. We call it a logarithmic box. Its appeal
is threefold. First, it is the minimal two-parameter shape (amplitude and width)
whose support is a clean interval, so it isolates the effect of boundaries from
the effect of curvature of the peak. Second, it interpolates controllably
between two physically distinct regimes: as $\Delta\to0$ it tends to the
monochromatic spectrum, while for $\Delta\gtrsim1$ its interior is locally
scale-invariant, with two well-separated edges. Third, the compactness of its
support makes every integral literally finite-domain, which lets us obtain the
GW spectrum in closed geometric terms rather than as Gaussian moments.

The box (top-hat) spectrum itself is not new. The GW signal from such a spectrum
was studied in Ref.~\cite{Saito:2009jt} for the first time, and we revisit this
case here in order to provide potentially useful methods and closed-form
expressions. This example was not among those treated in the semianalytic study~\cite{Kohri:2018awv}, so our detailed treatment constitutes a useful addition to the existing catalogue; a top-hat spectrum has also been recently
studied~\cite{Terada:2025cto}. The box is, moreover, of direct
practical relevance to data analysis: it is used in the pulsar-timing-array
analysis by the NANOGrav collaboration~\cite{NANOGrav:2023hvm}, while the
approximation of a generic curvature spectrum by a superposition of several
top-hat functions is used in Refs.~\cite{LISACosmologyWorkingGroup:2025vdz,LISACosmologyWorkingGroup:2024hsc} to
study the prospects of LISA to reconstruct the SIGW models.

This paper is organized as follows. In Sec.~\ref{sec:master} we review the
second-order tensor equation of motion, solve it during radiation domination,
and reduce the induced-GW energy density to the standard double convolution with
the time-averaged kernel, recording also the conversion to the present-day
density. In Sec.~\ref{sec:box} we define the logarithmic box, fix its
normalisation, and recast the master convolution as an integral over the compact
intersection of the box support with the momentum triangle, obtaining the exact
starting point and the hard ultraviolet endpoint. Section~\ref{sec:narrow}
treats the narrow box: we recover the delta limit, derive the geometric overlap
factor, extract the infrared break and the two power laws, estimate the
regularised resonance height, and confront the resulting template both with
direct numerics and with the narrow lognormal peak. Section~\ref{sec:broad}
treats the broad box: we introduce the two edge functions, expand the kernel at
the lower and upper edges, identify the scale-invariant plateau, assemble a
uniform all-scale formula together with the spectral maximum, and again compare
with numerics and with the broad lognormal peak. We summarise our results and
discuss extensions in Sec.~\ref{Con}. A closed-form, integral-free surrogate for
the edge functions, suitable for massive parameter scans, is collected in
Appendix~\ref{app:closed}.

%%%%%%%%%%%%%%%%%%%%%%%%%%%%%%%%%%%%%%%%%%%%%%%%%%%%%%%%%%%%%%%%%
\section{Scalar-induced gravitational waves in the radiation era}
\label{sec:master}
%%%%%%%%%%%%%%%%%%%%%%%%%%%%%%%%%%%%%%%%%%%%%%%%%%%%%%%%%%%%%%%%%
We adopt the conformal Newtonian gauge and, neglecting anisotropic
stress so that the two Bardeen potentials coincide up to a sign, write the
perturbed line element with a transverse-traceless tensor mode $h_{ij}$ as
\begin{equation}
  \dd s^2=a^2(\eta)\Big[-(1+2\Psi)\,\dd\eta^2
  +\big((1-2\Psi)\delta_{ij}+\tfrac12 h_{ij}\big)\dd x^i\dd x^j\Big],
  \label{eq:metric}
\end{equation}
where $\eta$ is conformal time and $\mathcal{H}\equiv a'/a$. {Here $\Psi$ is first-order primordial scalar perturbation and $h_{ij}$ is the second-order tensor perturbation. And we neglect the first-order tensor perturbation.} {We follow the radiation-era conventions of Refs.~\cite{Ananda:2006af,Baumann:2007zm,Kohri:2018awv,Espinosa:2018eve}.} Expanding the
spatial tensor part in momentum space and in the two transverse polarizations
$\lambda=+,\times$,
\begin{equation}
  h_{ij}(\eta,\bm x)=\int\frac{\dd^3k}{(2\pi)^{3/2}}
  \sum_{\lambda}\,\mathrm{e}^{\lambda}_{ij}(\hat{\bm k})\,
  h^\lambda_{\bm k}(\eta)\,e^{i\bm k\cdot\bm x},
\end{equation}
the quadratic part of the Einstein equations gives, for each polarization,
\begin{equation}
  h^{\lambda\,\prime\prime}_{\bm k}+2\mathcal{H}\,h^{\lambda\,\prime}_{\bm k}
  +k^2 h^\lambda_{\bm k}
  =4\,S^\lambda_{\bm k}(\eta),
  \label{eq:eom}
\end{equation}
with a source built bilinearly from the scalar potential,
\begin{equation}
  S^\lambda_{\bm k}=\int\frac{\dd^3 q}{(2\pi)^{3/2}}\,
  \mathrm{e}^\lambda_{ij}(\hat{\bm k})\,q^i q^j\,
  f(\bm k,\bm q,\eta)\,\Psi_{\bm q}\Psi_{\bm k-\bm q}.
  \label{eq:source}
\end{equation}
The polarization tensors project out the part of $q^iq^j$ transverse to
$\hat{\bm k}$. The time dependence is carried by the transfer function
$T(k\eta)$ defined through $\Psi_{\bm k}(\eta)=T(k\eta)\,\psi_{\bm k}$, with
$\psi_{\bm k}$ the primordial value, and the kernel $f$ collects the products of
transfer functions and their derivatives,
\begin{equation}
  f(\bm k,\bm q,\eta)=
  2\,T(q\eta)T(|\bm k-\bm q|\eta)
  +\tfrac{4}{3(1+w)}\!\left(\mathcal{H}^{-1}T'(q\eta)+T(q\eta)\right)
  \!\left(\mathcal{H}^{-1}T'(|\bm k-\bm q|\eta)+T(|\bm k-\bm q|\eta)\right),
  \label{eq:fkernel}
\end{equation}
where $w$ is the background equation of state. During radiation domination
$w=1/3$, $a\propto\eta$, $\mathcal{H}=1/\eta$, and the potential transfer
function for modes that re-enter the horizon is
\begin{equation}
  T(k\eta)=\frac{9}{(k\eta)^2}
  \!\left[\frac{\sqrt3}{k\eta}\sin\!\frac{k\eta}{\sqrt3}
  -\cos\!\frac{k\eta}{\sqrt3}\right].
  \label{eq:transfer}
\end{equation}

Equation~\eqref{eq:eom} is solved with the retarded Green function of the
operator on its left-hand side. In radiation domination the homogeneous
solutions are $\sin(k\eta)/k$ and $\cos(k\eta)/k$, and
\begin{equation}
  a(\eta)\,h^\lambda_{\bm k}(\eta)=
  \frac{4}{k}\int_{\eta_0}^{\eta}\dd\bar\eta\,
  a(\bar\eta)\,\sin\!\big(k(\eta-\bar\eta)\big)\,S^\lambda_{\bm k}(\bar\eta).
  \label{eq:green}
\end{equation}
The induced GW energy density per logarithmic wavenumber, normalized to the
critical density, is
\begin{equation}
  \OGW(\eta,k)=\frac{1}{{48}}\!\left(\frac{k}{\mathcal{H}}\right)^{\!2}
  \overline{\mathcal{P}_h(\eta,k)},
  \label{eq:omega-def}
\end{equation}
where $\mathcal{P}_h$ is the dimensionless tensor power spectrum defined from
the equal-time two-point function and the overbar denotes the average over the
fast oscillations of the sub-horizon modes. Squaring~\eqref{eq:green},
contracting the polarization structure, and writing the scalar two-point
function in terms of the curvature power spectrum $\PR$ (with
$\mathcal{P}_\psi=\tfrac49\PR$ during radiation domination), the angular
integrals can be done once and for all. Introducing the dimensionless ratios
\begin{equation}
  u=\frac{|\bm k-\bm q|}{k},\qquad v=\frac{q}{k},
\end{equation}
and carrying out the late-time average of the squared kernel as in
refs.~\cite{Kohri:2018awv,Espinosa:2018eve,Pi:2020otn}, one arrives at a
time-independent spectrum for sub-horizon scales,
\begin{equation}
  \OGW(k)=\int_0^\infty\!\dd v\int_{|1-v|}^{1+v}\!\dd u\;
  T(u,v)\,\PR(ku)\,\PR(kv).
  \label{eq:master}
\end{equation}
The kernel $T(u,v)$ is the product of the squared polarization factor, the oscillation average of the squared Green integral, and the phase-space Jacobian. Its
explicit radiation-era form is
\begin{align}
  T(u,v)=&\;\frac{3}{1024\,u^8 v^8}\,
  \big[4v^2-(1+v^2-u^2)^2\big]^2\,(u^2+v^2-3)^2
  \nonumber\\
  &\times
  \bigg\{\!\Big[(u^2+v^2-3)
  \ln\!\Big|\frac{3-(u+v)^2}{3-(u-v)^2}\Big|-4uv\Big]^2
  +\pi^2(u^2+v^2-3)^2\,\Heavi(u+v-\sqrt3)\bigg\}.
  \label{eq:kernel}
\end{align}
The Heaviside term is the imprint of the resonance between the tensor mode and
the scalar source travelling at sound speed $1/\sqrt3$: it switches on when the
two scalar momenta can add up to the resonant value, $u+v=\sqrt3$. The
logarithm inside the curly brackets diverges on the same surface, producing the
sharp resonant feature familiar from the monochromatic case. For later use we
note the relation to the kernel $\mathcal{T}(u,v)$ of ref.~\cite{Pi:2020otn},
\begin{equation}
  T(u,v)=\frac{3\,\mathcal{T}(u,v)}{u^2 v^2},
  \label{eq:T-relation}
\end{equation}
so that eq.~\eqref{eq:master} coincides with the master integral used there;
the rearrangement~\eqref{eq:kernel} is more convenient for the box because the
power of $uv$ is collected in a single prefactor. In the limit
$u\simeq v\to\infty$ the kernel decays as $T\sim u^{-4}\,[\ln(u+v)]^2$, a fact
that controls all the IR tails below. Equation~\eqref{eq:master} is the spectrum frozen in at the end of radiation
domination. Because the GW energy density redshifts as radiation, the value
measured today follows by the standard transfer of the radiation density and the
change in the effective number of relativistic species,
\begin{equation}
  \Omega_{\rm GW,0}(f)\,h^2=1.6\times10^{-5}
  \!\left(\frac{g_{*s}}{106.75}\right)^{\!-1/3}
  \!\left(\frac{\Omega_{r,0}h^2}{4.1\times10^{-5}}\right)\OGW(f),
  \label{eq:today}
\end{equation}
where the comoving wavenumber maps onto frequency through
$f=k/(2\pi a_0)\simeq 1.5\times10^{-9}\,(k/1\,{\rm pc}^{-1})\,$Hz.
{Here $g_*$ and $g_{*s}$ count the effective relativistic degrees of freedom in the energy and entropy densities, respectively. The normalization $\Omega_{r,0}h^2\simeq4.1\times10^{-5}$ includes photons and the relativistic-neutrino contribution \cite{Planck:2018vyg}. The  value $g_{*s}=g_*=106.75$ assumes the Standard Model plasma well above the electroweak scale \cite{Watanabe:2006qe,Saikawa:2018rcs}.}
In what follows we work with the radiation-era spectrum~\eqref{eq:master}; the present
amplitude is recovered by the constant factor in eq.~\eqref{eq:today}.

%%%%%%%%%%%%%%%%%%%%%%%%%%%%%%%%%%%%%%%%%%%%%%%%%%%%%%%%%%%%%%%%%
\section{The logarithmic box}
\label{sec:box}

We take the curvature power spectrum to be uniform in $\ln k$ over an interval
centered on $k_*$,
\begin{equation}
  \PR(k)=\frac{\AR}{2\Delta}\,
  \Heavi\!\left(\Delta-\Big|\ln\frac{k}{k_*}\Big|\right),
  \qquad k_\pm \equiv k_*\,e^{\pm\Delta}.
  \label{eq:box-spectrum}
\end{equation}
The two free parameters are the integrated amplitude $\AR$ and the
dimensionless half-width $\Delta$. {We use sharp edges as an idealization, rather than as an exact inflationary prediction. Approximately flat enhanced bands can arise through parametric amplification~\cite{Cai:2019bmk} or in curvaton scenarios~\cite{Ferrante:2023bgz}.} The normalization is fixed by
\begin{equation}
  \int_0^\infty\PR(k)\,\dd\ln k
  =\frac{\AR}{2\Delta}\int_{-\Delta}^{\Delta}\dd\xi=\AR,
  \qquad \xi\equiv\ln\frac{k}{k_*},
  \label{eq:norm}
\end{equation}
so that $\AR/(2\Delta)$ is the box height. Two limits are immediate. As
$\Delta\to0$ at fixed $\AR$, the normalized rectangle of width $2\Delta$ and
height $\AR/(2\Delta)$ becomes a Dirac mass,
$\PR(k)\to\AR\,\delta(\ln k/k_*)$, reproducing the monochromatic source. In the
opposite regime, any tensor mode whose support lies well inside $[k_-,k_+]$ sees
a locally constant  source of amplitude $\AR/(2\Delta)$; the
broad box is therefore a finite, sharply truncated piece of a scale-invariant
spectrum.

It is convenient to symmetrize the two scalar momenta in
eq.~\eqref{eq:master}. With
\begin{equation}
  \kappa\equiv\frac{k}{k_*},\qquad
  x=\frac{u+v}{2},\qquad y=\frac{u-v}{2},
  \label{eq:xy}
\end{equation}
one has $u=x+y$, $v=x-y$, and the Jacobian $\dd u\,\dd v=2\,\dd x\,\dd y$. The
momentum triangle $|1-v|\le u\le 1+v$ together with $v>0$ is equivalent to the
two inequalities $u+v\ge1$ and $|u-v|\le1$, i.e.
\begin{equation}
  x\ge\frac12,\qquad -\frac12\le y\le\frac12,
  \label{eq:triangle}
\end{equation}
a semi-infinite strip in the $(x,y)$ plane. A useful property of these variables
is that the leading prefactor of the kernel factorises,
\begin{equation}
  4v^2-(1+v^2-u^2)^2=(4x^2-1)(1-4y^2),
  \label{eq:factorisation}
\end{equation}
so that $T(u,v)$ vanishes on the three lines $x=1/2$ and $y=\pm1/2$ that bound
the strip; this feature will control the behaviour at the hard upper edge in
Sec.~\ref{sec:broad-uv}. The box constraint $\PR(ku)\PR(kv)\ne0$ requires
$k_-\le ku\le k_+$ and $k_-\le kv\le k_+$, which in the present variables reads
\begin{equation}
  a\le x+y\le b,\qquad a\le x-y\le b,
  \qquad a\equiv\frac{e^{-\Delta}}{\kappa},\quad
  b\equiv\frac{e^{\Delta}}{\kappa}.
  \label{eq:boxcond}
\end{equation}
For a given $x$ the four conditions in eq.~\eqref{eq:boxcond} bound $y$ from
below and above by
\begin{equation}
  y_-(x; {\kappa})=\max\Big[-\tfrac12,\;a-x,\;x-b\Big],\qquad
  y_+(x; {\kappa})=\min\Big[\tfrac12,\;b-x,\;x-a\Big].
  \label{eq:ylimits}
\end{equation}
The admissible range of $x$ is
\begin{equation}
  \max\!\Big(\tfrac12,\,a\Big)\le x\le b,
  \label{eq:xsupport}
\end{equation}
and the inner integral is empty whenever $y_+\le y_-$. The exact box convolution
is therefore the compact integral
\begin{equation}
  {\;
  \OGW^{\rm box}(\kappa)=\Big(\frac{\AR}{2\Delta}\Big)^2
  \int_{\max(1/2,a{(\kappa)})}^{b{(\kappa)}}\!\dd x
  \int_{y_-(x; {\kappa})}^{y_+(x; {\kappa})}2\,\dd y\;T(x+y,x-y).\;}
  \label{eq:exact-box}
\end{equation}
This is the common starting point for both the asymptotic analysis and the
numerical checks. {At fixed $\Delta$, the $\kappa$ dependence enters through $a(\kappa)$, $b(\kappa)$ and hence the integration limits $y_\pm(x;\kappa)$.} Two structural features follow at once. The lower limit
$x\ge a=e^{-\Delta}/\kappa$ pushes the integration region to large $x$ when
$\kappa$ is small, which is the origin of the IR tail. The upper limit $x\le b$
combined with $x\ge\tfrac12$ requires $b\ge\tfrac12$, hence
\begin{equation}
  \OGW^{\rm box}(\kappa)=0\qquad\text{for}\qquad \kappa>2\,e^{\Delta},
  \label{eq:hard-endpoint}
\end{equation}
which means no GW power is generated above twice the largest scalar
wavenumber, a direct consequence of the bounded support.

\section{Narrow box}
\label{sec:narrow}

We first treat $\Delta\ll1$, for which the entire scalar support is concentrated
near $k_*$. A narrow peak of this kind is a common prediction of PBH-oriented
inflationary models, including double-inflation and curvaton
scenarios~\cite{Kawasaki:1997ju,Frampton:2010sw,Kawasaki:2012wr,Ando:2017veq,Ando:2018qdb, Pi:2017gih,Inomata:2017okj}, and the
sound-speed-resonance
mechanism~\cite{Cai:2018tuh,Cai:2019jah,Chen:2019zza,Chen:2020uhe}, all of which
enhance $\PR$ over a narrow band $\Delta\ll1$, so that the box reduces to a
small perturbation of a monochromatic source.

\subsection{The delta limit and the resonance}
\label{sec:delta}

When $\Delta\to0$ the two box factors in eq.~\eqref{eq:master} become Dirac
masses in $\ln$, $\PR(q)\to\AR\,\delta(\ln q/k_*)$. Writing
$\xi_1=\ln(ku/k_*)$ and $\xi_2=\ln(kv/k_*)$, so that $\dd u=u\,\dd\xi_1$ and
$\dd v=v\,\dd\xi_2$, the two delta functions set $\xi_1=\xi_2=0$, i.e.
$u=v=1/\kappa$. The Jacobian from $(u,v)$ to $(\xi_1,\xi_2)$ supplies a factor
$uv$ evaluated at that point, and one obtains
\begin{equation}
  \frac{\OGWx{\delta}(\kappa)}{\AR^2}
  =\frac{1}{\kappa^2}\,T\!\left(\frac1\kappa,\frac1\kappa\right)\,
  \Heavi(2-\kappa).
  \label{eq:delta-master}
\end{equation}
The Heaviside reflects the requirement that the point $u=v=1/\kappa$ lie inside
the momentum triangle, which holds only for $\kappa\le2$. Substituting the
diagonal kernel from eq.~\eqref{eq:kernel}, with
$4v^2-(1+v^2-u^2)^2\big|_{u=v}=4u^2-1$, $(u^2+v^2-3)\big|_{u=v}=2u^2-3$ and
$\ln|(3-4u^2)/3|$ at $u=1/\kappa$, gives the closed form
\begin{align}
  \frac{\OGWx{\delta}(\kappa)}{\AR^2}
  =&\;\frac{3}{1024}\,\kappa^2(4-\kappa^2)^2(2-3\kappa^2)^2
  \nonumber\\
  &\times\!\left[\Big(\!-4+(2-3\kappa^2)\ln\Big|1-\frac{4}{3\kappa^2}\Big|\Big)^2
  +\pi^2(2-3\kappa^2)^2\,\Heavi\!\Big(\frac{2}{\sqrt3}-\kappa\Big)\right]\!\Heavi(2-\kappa).
  \label{eq:delta}
\end{align}
The logarithm diverges at $\kappa=2/\sqrt3$, where the resonance condition is met
on the diagonal; this is the familiar resonant peak of the monochromatic source.
Expanding eq.~\eqref{eq:delta} for $\kappa\ll1$, where
$\ln|1-4/(3\kappa^2)|\to -2\ln\kappa$, the prefactors approach
$\tfrac{3}{1024}\,\kappa^2\cdot16\cdot4$ and the bracket approaches
$16\ln^2\kappa+4\pi^2$, so that
\begin{equation}
  \frac{\OGWx{\delta}(\kappa)}{\AR^2}\xrightarrow[\kappa\ll1]{}
  3\,\kappa^2\ln^2\kappa,
  \label{eq:delta-IR}
\end{equation}
the universal $k^2\ln^2 k$ tail of a monochromatic source.

\subsection{The geometric overlap factor}
\label{sec:overlap}

\begin{figure}[t]
  \centering
  \includegraphics[width=0.43\textwidth]{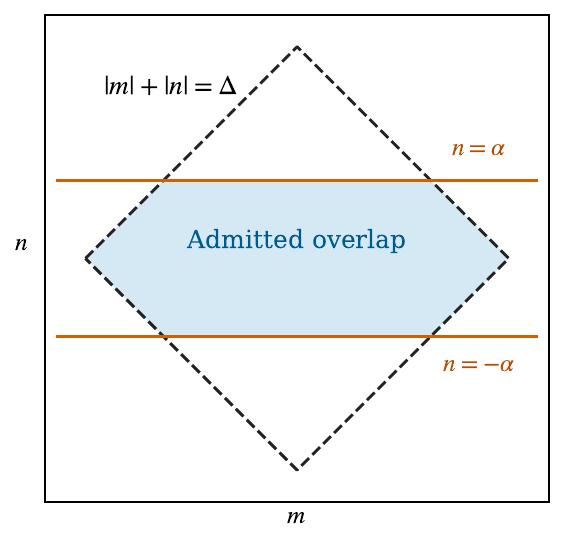}
  \caption{{The logarithmic box $|m|+|n|\le\Delta$ (dashed diamond) and the leading narrow-box triangle constraint $|n|\le\alpha$ (orange boundaries), for $0<\alpha<\Delta$. The blue region is their admitted overlap. The exact triangle boundaries are those in Eq.~\eqref{eq:triangle-mn}; the drawing represents their leading strip approximation.}}
  \label{fig:mn-domain}
\end{figure}

We now keep $\Delta$ small but nonzero and quantify the leading correction to
eq.~\eqref{eq:delta-master}. Parametrize the two scalar momenta in the box by
their log-displacements,
\begin{equation}
  ku=k_*\,e^{\xi_1},\qquad kv=k_*\,e^{\xi_2},\qquad
  \xi_1,\xi_2\in[-\Delta,\Delta],
\end{equation}
so that $u=e^{\xi_1}/\kappa$ and $v=e^{\xi_2}/\kappa$. It is natural to use the
mean and half-difference of the two displacements,
\begin{equation}
  m=\frac{\xi_1+\xi_2}{2},\qquad n=\frac{\xi_1-\xi_2}{2},
  \qquad u=\frac{e^{m+n}}{\kappa},\quad v=\frac{e^{m-n}}{\kappa},
  \label{eq:mn}
\end{equation}
with $\dd\ln u\,\dd\ln v=\dd\xi_1\,\dd\xi_2=2\,\dd m\,\dd n$. The square
$\xi_1,\xi_2\in[-\Delta,\Delta]^2$ is the set $|m+n|\le\Delta$ and
$|m-n|\le\Delta$, i.e. a diamond
\begin{equation}
  |m|+|n|\le\Delta,
  \label{eq:diamond}
\end{equation}
of area $2\Delta^2$ in the $(m,n)$ plane. {Figure~\ref{fig:mn-domain} shows the diamond and the leading strip approximation used below.} Rewriting the master integral in
log-momenta,
\begin{equation}
  \OGW^{\rm box}(\kappa)
  =\Big(\frac{\AR}{2\Delta}\Big)^2
  \int 2\,\dd m\,\dd n\;u v\,T(u,v),
  \label{eq:mn-master}
\end{equation}
where $uv\,T(u,v)$ is the smooth, non-singular combination
$\dd u\,\dd v\,T=\dd\ln u\,\dd\ln v\,(uvT)$, integrated over the intersection of
the diamond~\eqref{eq:diamond} with the momentum triangle. {Both $u=e^{m+n}/\kappa$ and $v=e^{m-n}/\kappa$ depend on $\kappa$.}

The key simplification is that, for a small diamond, the triangle constraints
are the only boundaries that can cut it, and they do so along smooth curves that
we may linearize. The two triangle conditions $u+v\ge1$ and $|u-v|\le1$ become,
using eq.~\eqref{eq:mn},
\begin{equation}
  \frac{2\,e^m\cosh n}{\kappa}\ge1,\qquad
  \frac{2\,e^m\,|\sinh n|}{\kappa}\le1.
  \label{eq:triangle-mn}
\end{equation}
For $0<\kappa<2$ the first condition is already satisfied at the diamond center
$m=n=0$, where its left-hand side equals $2/\kappa>1$, and it remains so
throughout an $O(\Delta)$ neighborhood; hence it is inactive at leading order.
The second condition is the binding one. Evaluating the slowly varying factor
$e^m$ at the center ($m=0$, an $O(\Delta)$ error in an already small region) it
reduces to $|\sinh n|\le\kappa/2$, i.e.
\begin{equation}
  |n|\le\alpha,\qquad \alpha\equiv\mathrm{arcsinh}\frac{\kappa}{2}.
  \label{eq:alpha}
\end{equation}
The finite width is thus felt, at leading order, purely as a geometric overlap:
the diamond $|m|+|n|\le\Delta$ must be intersected with the horizontal strip
$|n|\le\alpha$. Slicing the diamond at fixed $n$ (where $m$ runs over a segment
of length $2(\Delta-|n|)$), the admitted area is
\begin{equation}
  A_{\rm ov}=
  \begin{cases}
  \displaystyle\int_{-\Delta}^{\Delta}2(\Delta-|n|)\,\dd n=2\Delta^2,
  & \alpha\ge\Delta,\\[10pt]
  \displaystyle\int_{-\alpha}^{\alpha}2(\Delta-|n|)\,\dd n=4\alpha\Delta-2\alpha^2,
  & \alpha<\Delta.
  \end{cases}
  \label{eq:Aov}
\end{equation}
In the first case the strip contains the whole diamond and the box is invisible;
in the second the strip removes the two outer caps with $|n|>\alpha$. Dividing
by the full diamond area defines the overlap fraction
\begin{equation}
  \Phi_{\rm box}(\kappa,\Delta)=\frac{A_{\rm ov}}{2\Delta^2}=
  \begin{cases}
  1, & \alpha\ge\Delta,\\[4pt]
  \dfrac{2\alpha}{\Delta}-\dfrac{\alpha^2}{\Delta^2}, & \alpha<\Delta.
  \end{cases}
  \label{eq:Phi}
\end{equation}
Replacing the smooth factor $uvT$ by its central value
$\kappa^{-2}T(1/\kappa,1/\kappa)$ in eq.~\eqref{eq:mn-master}, multiplying by
$A_{\rm ov}$, and comparing with eq.~\eqref{eq:delta-master}, the constant
factor $(\AR/2\Delta)^2$ combines with the area so that the narrow-box spectrum
is exactly the delta result rescaled by the overlap fraction,
\begin{equation}
  {\;
  \OGW^{(\rm box,\,\Delta\ll1)}(\kappa)\simeq
  \Phi_{\rm box}(\kappa,\Delta)\,\OGWx{\delta}(\kappa).\;}
  \label{eq:narrow-phi}
\end{equation}
This is the box counterpart of the error-function factor that multiplies the
delta result for a narrow lognormal peak~\cite{Pi:2020otn}. Two remarks are in
order. First, the overlap factor is purely kinematic: it follows from how much
of the scalar log-momentum phase space the momentum triangle admits, and is
therefore independent of the explicit form of the kernel and, in fact, of the
background equation of state. {Second, the central-value replacement is controlled as $\Delta\to0$ at fixed $\kappa$ away from the resonance, zeros of the Dirac spectrum and its upper endpoint. It is not uniform when $\kappa\lesssim\Delta$: there the angular factor varies across the admitted strip. The overlap still identifies the infrared crossover, but its amplitude must be checked separately.}

\subsection{Infrared break and power laws}
\label{sec:break}

The overlap fraction changes regime when $\alpha=\Delta$, i.e. at
\begin{equation}
  \kappa_b=2\sinh\Delta\simeq2\Delta,
  \label{eq:kb}
\end{equation}
or in physical units $k_b=2k_*\sinh\Delta\simeq2k_*\Delta$. For
$\kappa\ll\kappa_b$ we have $\alpha\simeq\kappa/2\ll\Delta$ and, keeping the
leading term in eq.~\eqref{eq:Phi}, $\Phi_{\rm box}\simeq\kappa/\Delta$. {For the amplitude, however, the angular factor $(1-4y^2)^2$ in the large-$x$ kernel must be retained. Its integral over $-1/2<y<1/2$ is $8/15$, giving}
\begin{equation}
  \frac{\OGW^{(\rm box,\,\Delta\ll1)}}{\AR^2}\simeq
  {\frac{8}{5\Delta}}\,\kappa^3\ln^2\kappa,
  \qquad \kappa\ll2\Delta\quad(\text{far IR}),
  \label{eq:narrow-farIR}
\end{equation}
while for $2\Delta\ll\kappa\ll1$ one has $\alpha\simeq\kappa/2>\Delta$,
$\Phi_{\rm box}=1$, and the delta scaling is restored,
\begin{equation}
  \frac{\OGW^{(\rm box,\,\Delta\ll1)}}{\AR^2}\simeq
  3\,\kappa^2\ln^2\kappa,
  \qquad 2\Delta\ll\kappa\ll1\quad(\text{near IR}).
  \label{eq:narrow-nearIR}
\end{equation}
Thus the IR slope breaks from $k^3\ln^2 k$ to $k^2\ln^2 k$ as $k$ increases
through $k_b$. The break is the spectral fingerprint of the finite support: a
tensor mode of wavenumber $k\ll k_b$ has too long a wavelength to resolve the
scalar band, so it couples only to a fraction $\propto k$ of the available
scalar phase space, steepening the universal $k^2$ tail of the monochromatic
limit into $k^3$. Equation~\eqref{eq:kb} is the box version of the relation
between the break and peak frequencies; {once a crossover convention is specified, comparing the break and the}
{resonant peak provides a characteristic estimate of the source width,
$\Delta\simeq\mathrm{arcsinh}(\kappa_b/2)\simeq \kappa_b/2$}.

\subsection{Regularization of the resonance}
\label{sec:res}

Near the resonance $\kappa=\kappa_p\equiv2/\sqrt3$ the geometric factor is
inactive, $\Phi_{\rm box}(\kappa_p,\Delta)=1$ for any
$\Delta<\mathrm{arcsinh}(1/\sqrt3)\simeq0.55$, so eq.~\eqref{eq:narrow-phi}
inherits the logarithmic divergence of $\OGWx{\delta}$. In the exact
finite-width convolution this divergence is integrable and the peak is finite.
To estimate its {value at $\kappa_p$} we isolate the singular part of eq.~\eqref{eq:delta}. {This follows the resonance-averaging approach of Ref.~\cite{Pi:2020otn}; that work uses a uniform logarithmic frequency bin, whereas here we average over the scalar box at fixed tensor wavenumber.}
Near $\kappa_p$ the prefactor reduces to
$\tfrac{3}{1024}\,\tfrac43\cdot\tfrac{64}{9}\cdot4=\tfrac19$, while
$2-3\kappa^2\to-2$, so
\begin{equation}
  \frac{\OGWx{\delta}}{\AR^2}\simeq
  \frac49\!\left[\Big(2+\ln\Big|1-\frac{4}{3\kappa^2}\Big|\Big)^2
  +\pi^2\,\Heavi(\kappa_p-\kappa)\right].
  \label{eq:res-near}
\end{equation}
Writing $\kappa=\kappa_p\,e^{\delta}$, one has
$1-4/(3\kappa^2)\simeq2\delta$ for small $\delta$, so the singular logarithm is
$\ln|2\delta|$. The resonance is active over the range of $\kappa$ for which the
surface $u+v=\sqrt3$ crosses the box, i.e. $|\delta|\lesssim\Delta$. {At fixed $\kappa=\kappa_p$, the leading logarithm is $\ln|2m|$. Integrating over $n$ in the diamond gives the normalized weight $(\Delta-|m|)/\Delta^2$, so averaging Eq.~\eqref{eq:res-near} gives}
\begin{align}
  \frac{\OGW^{\rm res}{(\kappa_p)}}{\AR^2}
  &\simeq\frac49\!\left[{\int_{-\Delta}^{\Delta}\!\frac{\Delta-|m|}{\Delta^2}}
  \big(2+{\ln|2m|}\big)^2{\dd m}+\frac{\pi^2}{2}\right]
  \nonumber\\
  &=\frac49\!\left[\big(\ln(2\Delta)+{\tfrac12}\big)^2+{\frac54}+\frac{\pi^2}{2}\right].
  \label{eq:res-peak}
\end{align}
{Here $\langle\ln|2m|\rangle=\ln(2\Delta)-3/2$ and $\operatorname{Var}(\ln|2m|)=5/4$ for the triangular weight.} Equation~\eqref{eq:res-peak} grows as
$\tfrac49\ln^2\Delta$ for $\Delta\to0$. {The leading logarithmic term agrees with the narrow lognormal result~\cite{Pi:2020otn}. And the subleading terms depend on the averaging weight:} the
divergence is controlled by the same integrable logarithm, whose smoothing
{depends on both the width and the profile beyond leading logarithmic order.} The compact geometric formula~\eqref{eq:narrow-phi} does not by itself
capture this local smoothing; eq.~\eqref{eq:res-peak} supplements it {at $\kappa_p$}. {At $\kappa_p$, the relative errors of this estimate are $0.062\%$, $1.25\%$, $3.16\%$, $10.9\%$ and $28.5\%$ for $\Delta=10^{-3},10^{-2},0.02,0.05,0.1$, respectively; it is a narrow-width asymptote rather than a formula for arbitrary width.}

\subsection{Numerical results and comparison}
\label{sec:narrow-num}

We now confront the narrow-box results with direct numerical integration of the
exact compact convolution~\eqref{eq:exact-box}. We stress that the numerical
curves play no role in defining the analytic formulae; they serve only to
display the accuracy of the geometric narrow-box law over a representative range
of widths.
{The numerical integral is split at changes of the support boundaries and at $x=\sqrt3/2$, with a variable transformation at the logarithmic resonance. The residual is $R=\Omega_{\rm app}/\Omega_{\rm num}-1$; the horizontal zero line denotes exact agreement. {As shown in Table \ref{tab-narrow},} on 101 sampled points in each of $0.3\le\kappa\le0.7$ and $1.4\le\kappa\le1.8$, its absolute value remains below $1\%$ for the tested widths $\Delta=10^{-3},10^{-2},0.02$, but reaches $5.3\%$ and $18.4\%$ for $\Delta=0.05$ and $0.1$, respectively. These are sampled maxima in windows that exclude the infrared crossover, the Dirac-spectrum dip, the resonance and the cutoff.}

\begin{table}
\centering
\caption{{Sampled accuracy of the geometric narrow-box formula, Eq.~\eqref{eq:narrow-phi}.
For each width we used 101 logarithmically spaced points in
$0.3\leq\kappa\leq0.7$ and 101 linearly spaced points in
$1.4\leq\kappa\leq1.8$. The quantity reported is
$\max|R|$ over these 202 points.}}
\label{tab-narrow}
\begin{tabular}{cc}
\hline
$\Delta$ & Maximum absolute relative error (\%) \\
\hline
0.001 & 0.002208 \\
0.01  & 0.2204   \\
0.02  & 0.8762   \\
0.05  & 5.254    \\
0.1   & 18.36    \\
\hline
\end{tabular}
\end{table}

\begin{figure}[t]
  \centering
  \includegraphics[width=0.99\textwidth]{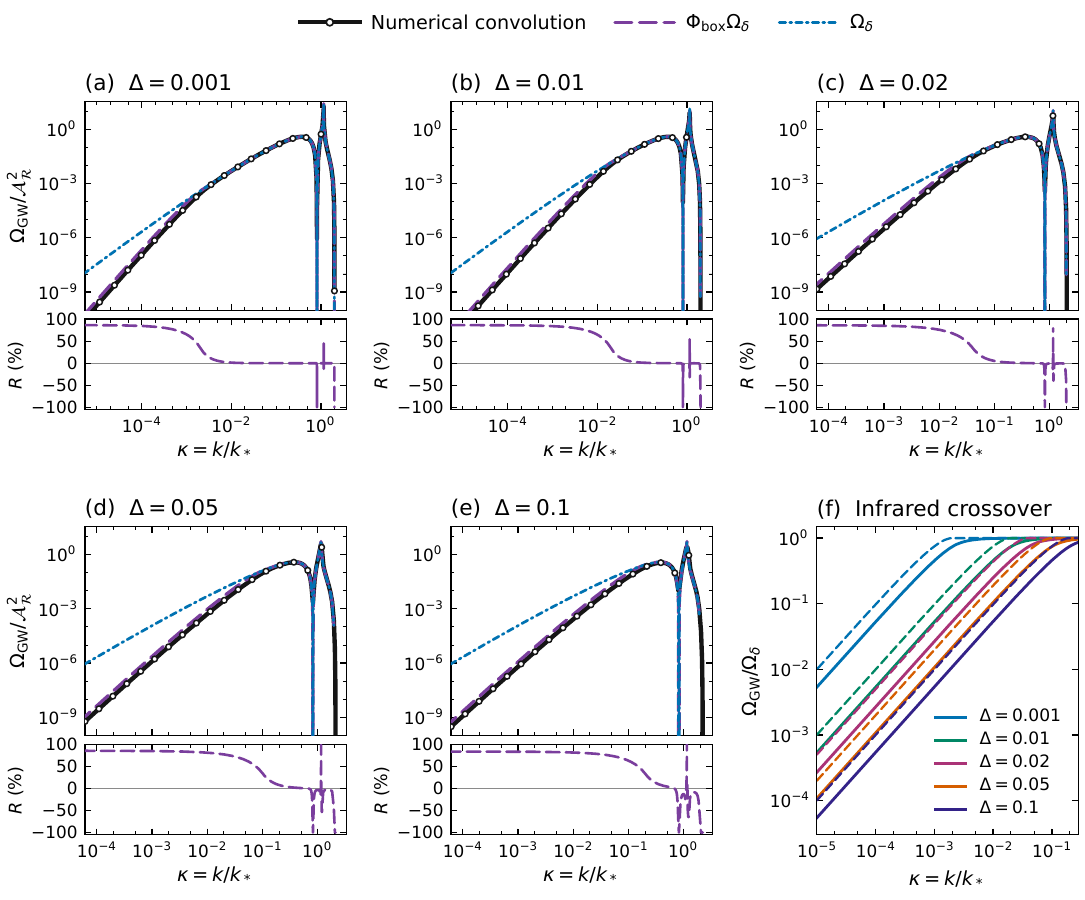}
  \caption{Narrow boxes with $\Delta=10^{-3},10^{-2},0.02,0.05,0.1$, normalised
  by $\AR^2$. {Black solid}: numerical integration of Eq.~\eqref{eq:exact-box}.
  {Violet dashed}: the geometric narrow-box formula \eqref{eq:narrow-phi}.
  {Blue dash-dotted}: the Dirac-spectrum result \eqref{eq:delta}.
  {The lower panels show $R=\Omega_{\rm app}/\Omega_{\rm num}-1$ for the geometric formula on linear ordinates. The last panel compares $\Omega_{\rm num}/\Omega_\delta$ (solid) and $\Phi_{\rm box}$ (dashed) across the five widths, making the infrared crossover visible.}
  The product factor $\Phi_{\rm box}$ bends the infrared tail from $k^2\ln^2k$
  down to $k^3\ln^2k$ below $\kappa_b=2\sinh\Delta$, {but overestimates its amplitude.}}
  \label{fig:narrow}
\end{figure}

Figure~\ref{fig:narrow} shows the narrow regime. As $\Delta$ decreases the
spectra approach the Dirac result over almost the whole range, departing from it
in the far infrared, where the overlap factor $\Phi_{\rm box}$ steepens the
slope from $k^2\ln^2k$ to $k^3\ln^2k$. {The scale $\kappa_b=2\sinh\Delta$ marks saturation of the geometric overlap, rather than a sharp break of the exact spectrum. Equation~\eqref{eq:narrow-phi} reproduces the two slopes, but overestimates the far-IR amplitude: in the ordered limit $\kappa/\Delta\to0$, $\Delta\to0$, the exact result is $8/15$ of the geometric estimate, so its relative error tends to $7/8$. Near $\kappa_p=2/\sqrt3$, the finite-width convolution also smooths the Dirac logarithmic divergence, as discussed in Sec.~\ref{sec:res}.}

It is instructive to place the narrow box alongside the lognormal peak of
Ref.~\cite{Pi:2020otn}. In the narrow limit both calculations amount to
averaging the Dirac answer over the small region of scalar momenta selected by
the box (or by the Gaussian weight) and then truncated by the momentum triangle.
For the lognormal the averaging weight is Gaussian and the truncated average
returns an error function; for the box the weight is flat and the truncation is
a literal area problem, the intersection of the diamond \eqref{eq:diamond} with
the strip $|n|\le\alpha$. The two play exactly the same role,
\begin{equation}
  \mathrm{erf}\!\left(\frac{\alpha}{\Delta}\right)
  \;\longleftrightarrow\;
  \Phi_{\rm box}(\kappa,\Delta)=
  \begin{cases}
  1, & \alpha\ge\Delta,\\
  2\alpha/\Delta-\alpha^2/\Delta^2, & \alpha<\Delta,
  \end{cases}
  \label{eq:erf-phi}
\end{equation}
and they share the same {powers in their} limiting behaviour: both tend to unity for
$\alpha\gg\Delta$ and grow linearly, $\propto\alpha\propto\kappa$, for
$\alpha\ll\Delta$. Consequently both shapes break from $k^3\ln^2k$ to
$k^2\ln^2k$ at {a comparable characteristic} scale $k_b\simeq2k_*\Delta$, which is set by
when the tensor wavelength becomes able to resolve the finite scalar support
rather than by any property of the weighting. Combined with the common
resonance location $\kappa_p=2/\sqrt3$, this yields the {characteristic-scale}
relation
\begin{equation}
  \frac{f_b}{f_p}\simeq\sqrt3\,\Delta ,
  \label{eq:fbfp}
\end{equation}
{with the same leading width scaling as the lognormal estimate~\cite{Pi:2020otn}. The numerical correspondence depends on the definition of width and of the crossover: $\Delta$ is the half-width for the box and the standard deviation in $\ln k$ for the lognormal.}

%%%%%%%%%%%%%%%%%%%%%%%%%%%%%%%%%%%%%%%%%%%%%%%%%%%%%%%%%%%%%%%%%
\section{Broad box}
\label{sec:broad}

In the opposite regime, $\Delta\gtrsim1$, the band is broad and the box interior is locally scale invariant. Such broad enhancements, extending over a wide range of scales, can arise, for example, through parametric amplification \cite{Cai:2019bmk} or in the curvaton scenario \cite{Ferrante:2023bgz}. For these spectra, the box provides a sharp-edged idealization of the resulting plateau.

\subsection{Edge variables}

When $\Delta\gtrsim1$ the band is wide and the source is no longer concentrated;
the integrand of Eq.~\eqref{eq:exact-box} explores regions far from the centre.
The natural variables are then the distances of the tensor mode from each edge,
\begin{equation}
  s=\frac{\kappa}{e^{-\Delta}}=\frac{k}{k_-},\qquad
  z=\frac{\kappa}{e^{+\Delta}}=\frac{k}{k_+},\qquad
  \omega(\kappa)=\frac{\OGW(\kappa)}{\AR^2},
  \label{eq:edge-vars}
\end{equation}
in terms of which $a=1/s$ and $b=1/z$.  We also use the rescaled spectrum
$\widehat\omega=(2\Delta)^2\omega$, normalised by the square of the box height,
while the physical observable remains $\OGW/\AR^2$.  In this notation
Eq.~\eqref{eq:exact-box} reads
\begin{equation}
  \widehat\omega_{\rm box}(s,z)=
  \int_{\max(1/2,\,1/s)}^{1/z}\!\dd x
  \int_{\max[-1/2,\,1/s-x,\,x-1/z]}^{\min[1/2,\,x-1/s,\,1/z-x]}
  \!2\,\dd y\;T(x+y,x-y),
  \label{eq:broad-exact}
\end{equation}
understood to vanish where the $y$-interval is empty.  The key structural point
is that for a broad band the two edges act independently: the lower edge is
governed by $s$ and the upper edge by $z$, with a common scale-invariant
interior between them.  We therefore define two universal, $\Delta$-independent
{edge functions},
\begin{align}
  \Em(s)&=
  \int_{\max(1/2,\,1/s)}^{\infty}\!\dd x
  \int_{\max[-1/2,\,1/s-x]}^{\min[1/2,\,x-1/s]}
  \!2\,\dd y\;T(x+y,x-y),
  \label{eq:edge-minus}\\
  \Ep(z)&=
  \int_{1/2}^{1/z}\!\dd x
  \int_{\max[-1/2,\,x-1/z]}^{\min[1/2,\,1/z-x]}
  \!2\,\dd y\;T(x+y,x-y),\qquad \Ep(z)=0\ \text{for}\ z\ge2.
  \label{eq:edge-plus}
\end{align}
$\Em(s)$ is the response of the kernel to a single lower scalar cut-off $q>k_-$
(no upper cut-off), and $\Ep(z)$ the response to a single upper cut-off $q<k_+$.
We now evaluate each region in turn.

\subsection{Lower edge}
\label{sec:broad-ir}

For $\kappa\ll e^{-\Delta}$, i.e. $s\ll1$, the support \eqref{eq:xsupport}
starts at $x=1/s\equiv r\gg1$, far out where $u,v\gg1$.  In this regime the
kernel admits a large-argument expansion.  Setting $u=x+y$, $v=x-y$ with
$x\gg1$ and $|y|\le1/2$, the building blocks of Eq.~\eqref{eq:kernel} behave as
\begin{equation}
  4v^2-(1+v^2-u^2)^2=(4x^2-1)(1-4y^2)\to4x^2(1-4y^2),\qquad
  u^2+v^2-3\to2x^2,
\end{equation}
\begin{equation}
  4uv\to4x^2,\qquad
  \ln\!\left|\frac{3-(u+v)^2}{3-(u-v)^2}\right|
  \to\ln\frac{4x^2}{3-4y^2}=\ln(4x^2)-\ln(3-4y^2),
\end{equation}
with $\Heavi(u+v-\sqrt3)=1$.  Collecting powers of $x$ (the prefactor scales as
$x^{-16}\cdot x^{4}\cdot x^{4}\cdot x^{4}=x^{-4}$) gives
\begin{equation}
  2T(x+y,x-y)\simeq
  \frac{3}{2x^4}(1-4y^2)^2
  \left[\big\{\ln(4x^2)-2-\ln(3-4y^2)\big\}^2+\pi^2\right].
  \label{eq:large-x-kernel}
\end{equation}
For $r\gg1$ the integral is dominated by $x\sim r$; the corner where the
$y$-range shrinks below $|y|\le1/2$ occupies an interval of width $\tfrac12$
near $x=r$ and contributes only at subleading order $O(r^{-4})$.  Taking
$|y|\le1/2$ throughout and extending the upper limit to infinity (the upper edge
is irrelevant here), the $y$-integral of Eq.~\eqref{eq:large-x-kernel} factorises
through the moments
\begin{equation}
  I_n=\int_{-1/2}^{1/2}\!\dd y\,(1-4y^2)^2\big[\ln(3-4y^2)\big]^n,
  \label{eq:In}
\end{equation}
which are elementary integrals; numerically
\begin{equation}
  I_0=\frac{8}{15},\qquad I_1=0.558953,\qquad I_2=0.587776 .
\end{equation}
Writing $L(x)=\ln(4x^2)-2$ and expanding the square,
\begin{equation}
  \int_{-1/2}^{1/2}\!\dd y\,2T
  =\frac{3}{2x^4}\Big[I_0\,L^2-2I_1\,L+\big(I_2+\pi^2I_0\big)\Big].
\end{equation}
The radial integrals $J_n=\int_r^\infty x^{-4}L^n\,\dd x$ follow by repeated
integration by parts, using $\dd L/\dd x=2/x$:
\begin{equation}
  J_0=\frac{r^{-3}}{3},\qquad
  J_1=\frac{r^{-3}}{3}\!\left(M_r+\frac23\right),\qquad
  J_2=\frac{r^{-3}}{3}\!\left(M_r^2+\frac43 M_r+\frac{8}{9}\right),
\end{equation}
with $M_r\equiv L(r)=\ln(4r^2)-2$.  Assembling
$\Em(s)\simeq\frac32[I_0J_2-2I_1J_1+(I_2+\pi^2I_0)J_0]$ and collecting powers of
$M_r$ gives the lower-edge spectrum
\begin{equation}
  \frac{\Omega^{\rm IR}_{\rm broad}}{\AR^2}
  \simeq\frac{\Em(s)}{(2\Delta)^2}
  =\frac{r^{-3}}{(2\Delta)^2}\big(c_0M_r^2+c_1M_r+c_2\big),
  \qquad M_r=\ln(4r^2)-2 ,
  \label{eq:broad-ir}
\end{equation}
with the coefficients
\begin{equation}
  c_0=\frac{I_0}{2}=\frac{4}{15},\qquad
  c_1=\frac{2I_0}{3}-I_1=-0.203398,\qquad
  c_2=\left(\frac49+\frac{\pi^2}{2}\right)I_0-\frac{2I_1}{3}+\frac{I_2}{2}=2.79018 .
  \label{eq:cs}
\end{equation}
Since $r=k_-/k$, Eq.~\eqref{eq:broad-ir} is the law
$\OGW\propto(k/k_-)^3\ln^2(k_-/k)$: the broad box rises in the far infrared with
the universal $k^3\ln^2k$ slope set entirely by the lower edge, modulated by the
slowly varying quadratic in $M_r$.

\subsection{Scale-invariant interior}
\label{sec:broad-mid}

When the tensor mode lies well inside the band, $k_-\ll k\ll k_+$, the kernel's
support never reaches either edge and the source is effectively the constant
$\PR=\AR/(2\Delta)$.  The convolution then collapses to the pure kernel integral,
\begin{equation}
  \frac{\Omega^{\rm mid}_{\rm broad}}{\AR^2}\simeq\frac{\Csi}{(2\Delta)^2},
  \qquad
  \Csi=\int_{1/2}^{\infty}\!\dd x\int_{-1/2}^{1/2}\!2\,\dd y\;T(x+y,x-y)
  \simeq0.82.
  \label{eq:plateau}
\end{equation}
The constant $\Csi$ is, by construction, the amplitude of the induced GW
spectrum sourced by an exactly scale-invariant curvature spectrum of unit
amplitude; its value is consistent with the scale-invariant limit of the
induced background discussed previously in the literature~\cite{Kohri:2018awv,Pi:2020otn}.
The broad box therefore develops a flat plateau of height $\Csi/(2\Delta)^2$
spanning the interior of the band.

\subsection{Upper edge}
\label{sec:broad-uv}

Near the hard endpoint \eqref{eq:hard-endpoint} the support shrinks to a small
neighbourhood of $u=v=1/2$.  Let
\begin{equation}
  \epsilon=\frac{e^{\Delta}}{\kappa}-\frac12=\frac1z-\frac12,\qquad 0<\epsilon\ll1,
  \label{eq:eps}
\end{equation}
so that $\kappa\to2 e^{\Delta}$ as $\epsilon\to0^+$.  Writing $x=\tfrac12+\delta$
with $0\le\delta\le\epsilon$, the limits \eqref{eq:ylimits} reduce (for a broad
box the lower cut-off $a=e^{-2\Delta}b\ll\tfrac12$ is inactive) to the small
triangle $|y|\le\epsilon-\delta$.  Two simplifications occur there.  First, the
resonance is absent, because $u+v=1+2\delta<\sqrt3$, so the $\pi^2$ term drops
out.  Second, the prefactor \eqref{eq:factorisation} vanishes quadratically:
$4v^2-(1+v^2-u^2)^2=(4x^2-1)(1-4y^2)\to4\delta(1-4y^2)$.  Expanding the kernel
about $u=v=\tfrac12$, where $u^2+v^2-3\to-\tfrac52$, $4uv\to1$ and
$\ln|(3-(u+v)^2)/(3-(u-v)^2)|\to\ln[2/(3-4y^2)]$, and evaluating the slowly
varying remainder at $y\to0$, one finds
\begin{equation}
  T\simeq\frac{3}{1024}\,4^8\cdot16\,\delta^2\cdot\frac{25}{4}\,
  \left(\frac52\ln\frac32-1\right)^2
  =19200\,\delta^2\left(\frac52\ln\frac32-1\right)^2 ,
\end{equation}
where we used $-\tfrac52\ln(2/3)-1=\tfrac52\ln\tfrac32-1$.  Integrating over the
small triangle, $\int_{-(\epsilon-\delta)}^{\epsilon-\delta}2\,\dd y=4(\epsilon-\delta)$
and $\int_0^\epsilon\delta^2(\epsilon-\delta)\,\dd\delta=\epsilon^4/12$, so
\begin{equation}
  \frac{\Omega^{\rm UV}_{\rm broad}}{\AR^2}
  \simeq\left(\frac{1}{2\Delta}\right)^2\!\frac{4\cdot19200}{12}
  \left(\frac52\ln\frac32-1\right)^2\!\epsilon^4
  =\frac{6400}{(2\Delta)^2}\left(\frac52\ln\frac32-1\right)^2\epsilon^4 .
  \label{eq:broad-uv}
\end{equation}
The induced spectrum thus approaches its hard endpoint as $\epsilon^4$, i.e. as
a quartic in the distance from $k=2k_+$, before vanishing identically beyond it.

\subsection{A uniform formula for all wavenumbers}

The three regimes above are the limits of a single object.  Writing the
normalised edge functions $g_\pm=\Em\!/\Csi,\ \Ep\!/\Csi$, which tend to unity in
the interior, the multiplicative composite of matched-asymptotics theory,
\begin{equation}
 {\;
  \omega_{\rm glob}(\kappa,\Delta)
  =\frac{1}{(2\Delta)^2}\,\frac{\Em(s)\,\Ep(z)}{\Csi},
  \qquad s=\kappa e^{\Delta},\ \ z=\kappa e^{-\Delta},
  \;}
  \label{eq:broad-global}
\end{equation}
reproduces each limit exactly.  Indeed, in the lower-edge regime $z\ll1$ so
$\Ep\to\Csi$ and $\omega_{\rm glob}\to\Em/(2\Delta)^2$, matching
Eq.~\eqref{eq:broad-ir}; in the interior $s\gg1$, $z\ll1$ so both edge functions
saturate and $\omega_{\rm glob}\to\Csi/(2\Delta)^2$, matching
Eq.~\eqref{eq:plateau}; and near the endpoint $s\gg1$ so $\Em\to\Csi$, while
$\Ep\to K\epsilon^4$ with
\begin{equation}
  K=6400\left(\frac52\ln\frac32-1\right)^2 ,
  \label{eq:K}
\end{equation}
reproducing the quartic cut-off \eqref{eq:broad-uv}.  Equation
\eqref{eq:broad-global} contains no fitted parameters: its only ingredients are
the two universal edge integrals and their common saturation value $\Csi$.  It is
exact in the three limits and provides a uniform interpolation everywhere in
between, becoming arbitrarily accurate as the band widens and the two edges
separate.  An integral-free closed form of $\Em$ and $\Ep$, convenient for large
scans, is given in Appendix~\ref{app:closed}.

For a broad box the spectrum is maximal on the plateau, so from
Eq.~\eqref{eq:plateau}
\begin{equation}
  \frac{\OGW^{\rm max}}{\AR^2}\simeq\frac{\Csi}{(2\Delta)^2}
  \simeq\frac{0.205}{\Delta^2}\qquad(\Delta\gtrsim1),
  \label{eq:broad-max}
\end{equation}
which depends only on the integrated power $\AR$ and the width.  For a narrow
box the largest value is reached at the resonance
$\kappa_p=2/\sqrt3$ rather than on a plateau, since there is no scale-invariant
interior; the amplitude there grows only as $\ln^2\Delta$ with decreasing $\Delta$,
as discussed in Sec.~\ref{sec:res}.  %Equation~\eqref{eq:broad-max} is the box analogue of the broad-peak maximum $\OGW^{\rm max}/\AR^2\simeq0.125/\Delta^2$ of the lognormal case~\cite{Pi:2020otn}; the difference in the numerical coefficient reflects the different relation between the integrated power and the peak height for a top-hat as opposed to a Gaussian profile.

\subsection{Numerical results and comparison}
\label{sec:broad-num}

We now turn to the broad regime, again comparing the analytic construction with
direct numerical integration of the exact compact convolution~\eqref{eq:exact-box}.

\begin{figure}[t]
  \centering
  \includegraphics[width=0.99\textwidth]{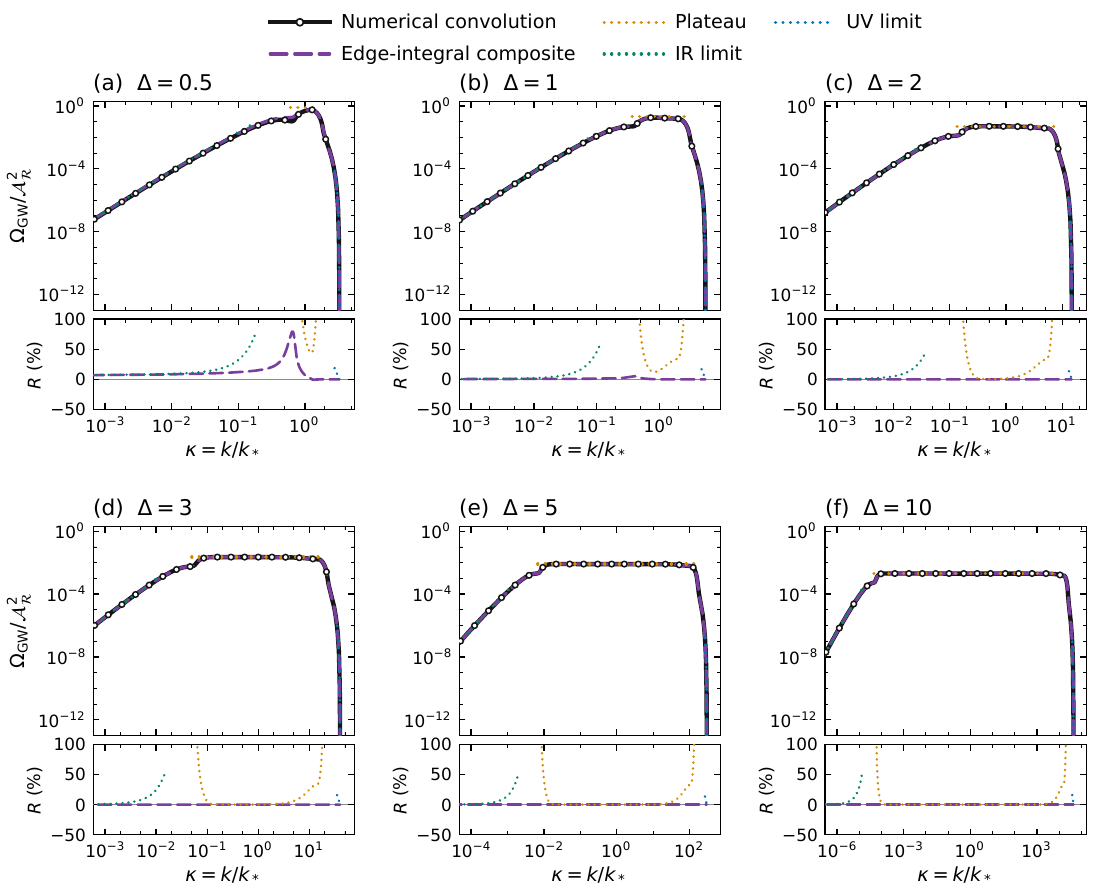}
  \caption{Broad boxes with $\Delta=0.5,1,2,3,{5,}10$, normalised by $\AR^2$.
  {Black solid: the compact numerical convolution. Violet dashed}:
  the uniform formula \eqref{eq:broad-global}, covering the full support
  $0<\kappa<2 e^{\Delta}$. {Orange, green and blue dotted}
  curves show the plateau \eqref{eq:plateau}, lower-edge \eqref{eq:broad-ir}, and
  upper-edge \eqref{eq:broad-uv} limits. {The lower panels show $R=\Omega_{\rm app}/\Omega_{\rm num}-1$ with linear ordinates; $R=0$ denotes agreement. Asymptotes appear only near their limits.}
}
  \label{fig:broad}
\end{figure}

Figure~\ref{fig:broad} shows the complementary broad regime. The lower edge
follows the universal $k^3\ln^2k$ rise of Eq.~\eqref{eq:broad-ir}; the interior
settles onto the flat plateau $\Csi/(2\Delta)^2$; and the upper edge plunges as
the quartic law \eqref{eq:broad-uv} before terminating at $\kappa=2e^{\Delta}$.
The composite \eqref{eq:broad-global} ({violet dashed}) joins these three behaviours into a
single curve over the entire support, with no free parameters. {The residual panels quantify the remaining differences, including the overlapping-edge regime at $\Delta=0.5$.} {{As shown in Table \ref{tab-broad},} on the sampled grids $10^{-3}e^{-\Delta}\le\kappa\le1.999e^{\Delta}$, where the maximum is taken on each grid, the largest absolute relative errors of the edge-integral composite are approximately $80\%$, $5.2\%$, $0.039\%$ and $2.1\times10^{-4}\%$ for $\Delta=0.5,1,2,3$, respectively. These are sampled maxima, not bounds for every wavenumber or intermediate width. At $\Delta=5,10$ the sampled differences are below $10^{-6}\%$.}

\begin{table}
\caption{{Sampled broad-box errors for the edge-integral composite,
Eq.~\eqref{eq:broad-global}. We used 581 logarithmically spaced points in $10^{-3}e^{-\Delta}\le\kappa\le1.999e^{\Delta}$. The quantity reported is $\max|R|$ over these 581 points.}
}
\label{tab-broad}
\centering
\begin{tabular}{cc}
\hline
$\Delta$ & Edge integral (\%) \\
\hline
0.5  & 79.56 \\
1    & 5.177 \\
1.25 & 1.551 \\
1.5  & 0.4627 \\
2    & 0.03868 \\
3    & 0.0002137 \\
5    & $<10^{-6}$ \\
10   & $<10^{-6}$ \\
\hline
\end{tabular}
\end{table}

The broad limit sharpens the contrast with the lognormal peak of
Ref.~\cite{Pi:2020otn}. A wide lognormal remains a smooth probability weight in
$\ln q$, and the induced background is a single smooth peak whose width is
inherited, reduced by $\sqrt2$, from the scalar profile. A wide box instead
possesses a scale-invariant interior bounded by two hard walls, so its
induced background is not a single smooth peak but a flat plateau capped by a
lower-edge rise and a quartic upper cut-off, encoded in the product of two edge
responses, Eq.~\eqref{eq:broad-global}.

%%%%%%%%%%%%%%%%%%%%%%%%%%%%%%%%%%%%%%%%%%%%%%%%%%%%%%%%%%%%%%%%%
\section{\label{Con}Summary and discussion}
We have computed the scalar-induced gravitational-wave background produced
during radiation domination by a box-shaped curvature power spectrum, a top-hat
of half-width $\Delta$ in $\ln k$. The sharp edges of the box turn the
finite-width physics into a transparent problem of geometry, and this allowed us
to obtain compact analytic results in both relevant regimes.

For a narrow box ($\Delta\ll1$) we {constructed a geometric approximation}
{given by} the Dirac-spectrum result times a purely
geometric overlap fraction $\Phi_{\rm box}(\kappa,\Delta)$,
Eqs.~\eqref{eq:Phi}--\eqref{eq:narrow-phi}. {The far-IR amplitude requires the angular weight in Eq.~\eqref{eq:narrow-farIR}, and the finite resonance requires the box average in Eq.~\eqref{eq:res-peak}.} Because $\Phi_{\rm box}$ arises from
the kinematic triangle alone, it is independent of the cosmological equation of
state, and it produces an infrared break from $k^3\ln^2k$ to $k^2\ln^2k$ at
$k_b=2k_*\sinh\Delta\simeq2k_*\Delta$. For a broad box ($\Delta\gtrsim1$) we
isolated three universal contributions, the lower-edge rise
$\OGW\propto(k/k_-)^3\ln^2(k_-/k)$ of Eq.~\eqref{eq:broad-ir}, the scale-invariant
plateau $\Csi/(2\Delta)^2$ of Eq.~\eqref{eq:plateau}, and the quartic upper
cut-off of Eq.~\eqref{eq:broad-uv} at the hard endpoint $k=2k_+$, and we combined
them into a single, parameter-free uniform formula \eqref{eq:broad-global}
written as a product of two $\Delta$-independent edge functions. A closed-form,
integral-free surrogate for these edge functions is given in
Appendix~\ref{app:closed}, reducing the broad-box template to elementary
operations suitable for massive parameter scans.

Comparing with the lognormal peak clarifies which features of an induced
background are dictated by the support of the scalar enhancement and which by
its detailed profile. The infrared break and the {leading width scaling}
$f_b/f_p\simeq\sqrt3\,\Delta$ are common to both shapes {with the width and crossover conventions specified above};
the plateau and the hard ultraviolet edge are peculiar to the box and signal a
finite band of nearly constant power. In this sense the box and the lognormal
bracket the two natural idealisations of a localised enhancement, sharp support
versus smooth tails, and a real spectrum will generically interpolate between
them.

Several extensions are natural. {The phase-space geometry underlying Eq.~\eqref{eq:narrow-phi} is independent of the equation of state. Applying the amplitude approximation to an early matter or kination era requires a separate check of the corresponding kernel}~\cite{Domenech:2019quo,Domenech:2020kqm,Inomata:2019ivs,Inomata:2019zqy,Papanikolaou:2020qtd}. A box
with soft rather than sharp edges, obtained by convolving
Eq.~\eqref{eq:box-spectrum} with a narrow window in $\ln k$, would round the
strict quartic endpoint and the resonance while leaving the lower-edge law and
the interior plateau intact, provided the smoothing scale is below $\Delta$.
Finally, the same edge-function bookkeeping should permit analytic templates for
piecewise-constant spectra built from several adjacent boxes, a flexible basis
for fitting measured or simulated backgrounds~\cite{LISACosmologyWorkingGroup:2025vdz}.
We leave these directions to future work.

\appendix

\section{Closed-form surrogate for the edge functions}
\label{app:closed}

The uniform formula \eqref{eq:broad-global} still contains the two universal
edge integrals $\Em(s)$ and $\Ep(z)$.  Although both are one-dimensional and
inexpensive, it is convenient for very large scans to replace them by explicit
elementary functions.  We do so through an ``opacity'' representation that
automatically respects the bounds $0\le\Em,\Ep\le\Csi$:
\begin{equation}
  \Em^{\rm cl}(s)=\Csi\big(1-e^{-Q_-(s)}\big),\qquad
  \Ep^{\rm cl}(z)=\Csi\big(1-e^{-Q_+(z)}\big).
  \label{eq:closed-opacity}
\end{equation}
When $Q_\pm\ll1$ the edge asymptotics are reproduced, $\Em^{\rm cl}\simeq\Csi Q_-$,
and when $Q_\pm\gg1$ the response saturates to the plateau $\Csi$.

\paragraph{Lower edge.}
Normalising Eq.~\eqref{eq:broad-ir} by $\Csi$ gives the leading opacity
\begin{equation}
  F_-(s)=\frac{s^3}{\Csi}\big[c_0L_s^2+c_1L_s+c_2\big],
  \qquad L_s=\ln\frac{4}{s^2}-2,
  \label{eq:Fminus}
\end{equation}
with $c_0,c_1,c_2$ from Eq.~\eqref{eq:cs} and $s=1/r$.  The full closed opacity,
which adds a smooth saturation as the mode enters the interior, is
\begin{equation}
  Q_-(s)=\frac{F_-(s)}{1+(s/s_-)^{\nu_-}}
  +A_-\ln\!\left[1+\left(\frac{s}{\bar s_-}\right)^{\mu_-}\right],
  \label{eq:Qminus}
\end{equation}
where $(s_-,\nu_-,A_-,\bar s_-,\mu_-)
  =(0.422419,\,2.727379,\,0.522543,\,1.347995,\,12.000000)$.
The first term preserves the $s^3\ln^2 s$ rise; the second supplies a rapid but
smooth approach to saturation.

\paragraph{Upper edge.}
With $\epsilon=1/z-1/2$ and the leading opacity
\begin{equation}
  F_+(z)=\frac{K}{\Csi}\,\epsilon^4,\qquad
  K=6400\left(\frac52\ln\frac32-1\right)^2,
  \label{eq:Fplus}
\end{equation}
from Eq.~\eqref{eq:K}, the full closed opacity is
\begin{equation}
  Q_+(z)=\frac{F_+(z)}{1+(\epsilon/\epsilon_+)^{\nu_+}}
  +A_+\ln\!\left[1+\left(\frac{\epsilon}{\bar\epsilon_+}\right)^{\mu_+}\right],
  \label{eq:Qplus}
\end{equation}
where $(\epsilon_+,\nu_+,A_+,\bar\epsilon_+,\mu_+)
  =(0.277569,\,1.614801,\,0.132339,\,0.421128,\,11.313723)$.
Because $\mu_+>4$, the logarithmic term is subleading as $\epsilon\to0$ and the
quartic endpoint $\Ep^{\rm cl}\simeq K\epsilon^4$ is preserved.

\paragraph{Integral-free template.}
Substituting Eqs.~\eqref{eq:closed-opacity}--\eqref{eq:Qplus} into the
uniform composite gives the completely integral-free broad-box spectrum
\begin{equation}
  {\;
  \omega_{\rm cl}(\kappa,\Delta)=
  \frac{1}{(2\Delta)^2}\,
  \frac{\Em^{\rm cl}(s)\,\Ep^{\rm cl}(z)}{\Csi},
  \qquad s=\kappa e^{\Delta},\ \ z=\kappa e^{-\Delta},
  \;}
  \label{eq:closed-global}
\end{equation}
set to zero outside $0<\kappa<2 e^{\Delta}$.  The numerical constants $(\epsilon_\pm,\nu_\pm,A_\pm,\bar\epsilon_\pm,\mu_\pm)$ are obtained by fitting
the two universal one-edge functions \eqref{eq:edge-minus}--\eqref{eq:edge-plus}
alone; no $\Delta$-dependent spectrum enters the fit.  

\begin{figure}[t]
  \centering
  \includegraphics[width=0.98\textwidth]{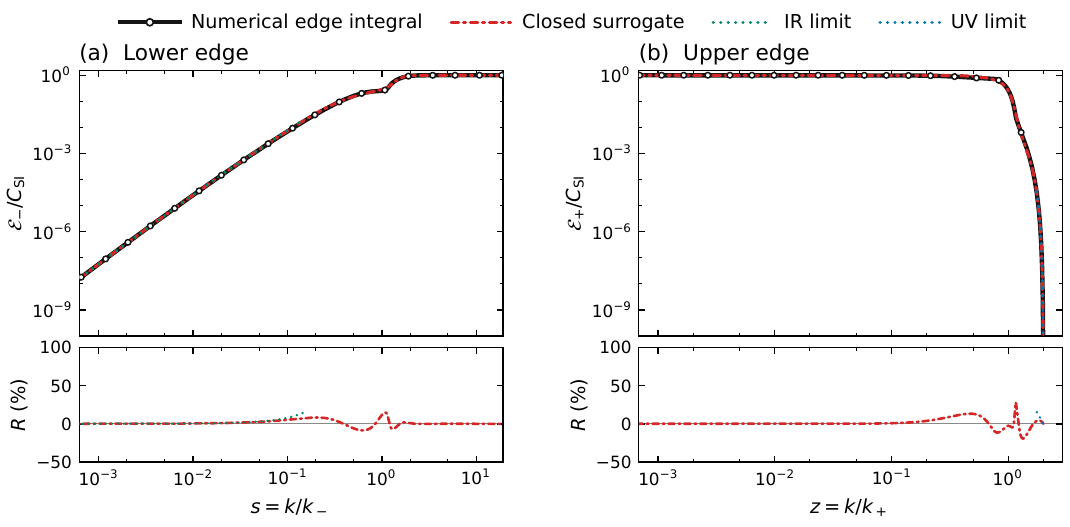}
  \caption{Universal lower- and upper-edge functions normalised by $\Csi$.
  {Black solid }: the edge integrals \eqref{eq:edge-minus} and \eqref{eq:edge-plus}.
  {Red dash-dotted}: the closed surrogate
  \eqref{eq:closed-opacity}--\eqref{eq:Qplus}. {Green and blue dotted}: the leading
  infrared and ultraviolet limits{, respectively}. {The lower panels show $R={\cal E}_{\rm app}/{\cal E}_{\rm num}-1$ against the corresponding edge integral on linear ordinates, including the transition regions. }}
  \label{fig:closed-edges}
\end{figure}

\begin{figure}[t]
  \centering
  \includegraphics[width=0.99\textwidth]{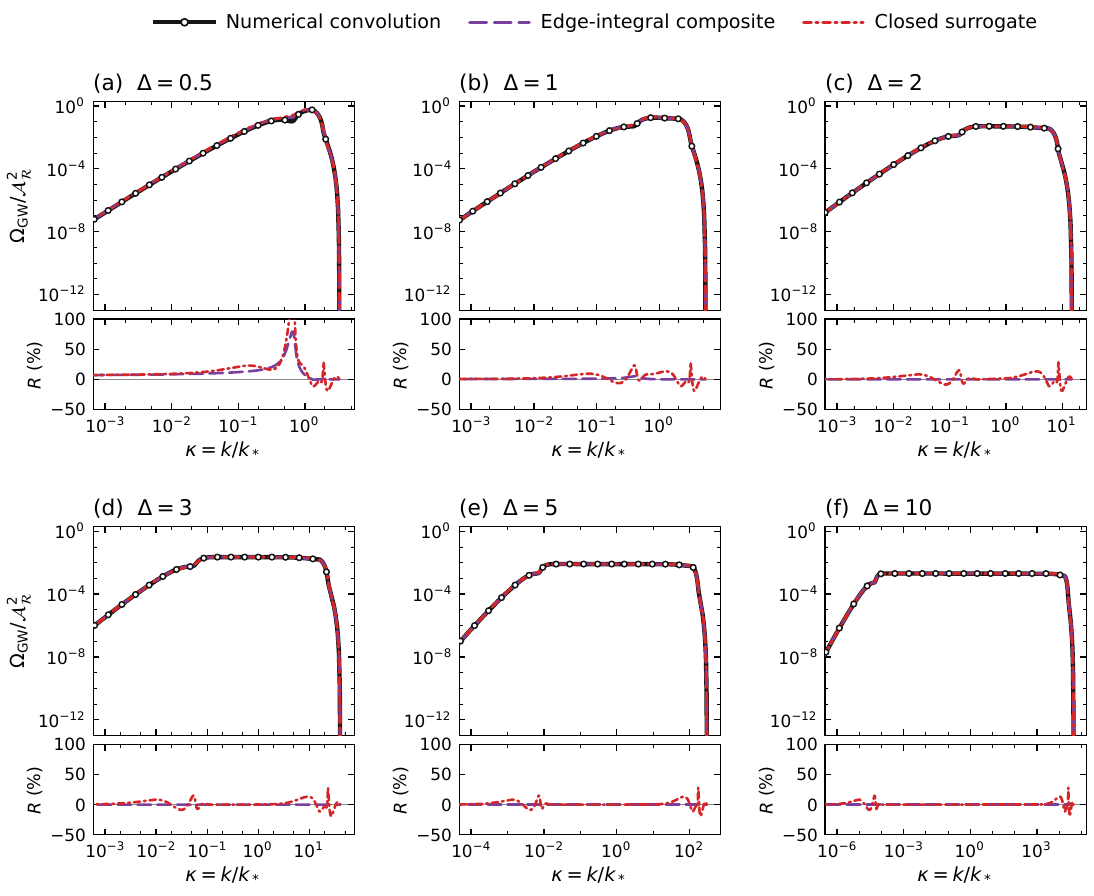}
  \caption{Broad-box spectra from the edge-integral composite
  \eqref{eq:broad-global} and from the integral-free closed form
  \eqref{eq:closed-global}, compared with numerical integration of
  Eq.~\eqref{eq:exact-box}{, for $\Delta=0.5,1,2,3,5,10$}. {Black solid : numerical convolution. Violet dashed: edge-integral composite. Red dash-dotted: closed form. The lower panels show $R=\Omega_{\rm app}/\Omega_{\rm num}-1$ on linear ordinates. }}
  \label{fig:closed-broad}
\end{figure}

Figures~\ref{fig:closed-edges} and \ref{fig:closed-broad} confirm the quality of
this construction.  Figure~\ref{fig:closed-edges} compares the two universal edge
functions with their closed-form opacities: the surrogate tracks the exact
integrals \eqref{eq:edge-minus}--\eqref{eq:edge-plus} over the entire range,
reproducing the $s^3\ln^2 s$ rise of the lower edge and the $\epsilon^4$ decay of
the upper edge while saturating smoothly to $\Csi$ in the interior.
Figure~\ref{fig:closed-broad} then feeds these surrogates into the composite
\eqref{eq:closed-global} and compares the result with direct numerical
integration of Eq.~\eqref{eq:exact-box}, for the same widths as in
Fig.~\ref{fig:broad}.  The integral-free form follows the full spectrum across
the entire support $0<\kappa<2e^{\Delta}$ for every width shown, with the largest
deviations confined to the narrow transition shoulders where the edge functions
cross over to the plateau; these residuals shrink as $\Delta$ grows and the two
edges separate, confirming that the closed template is amply accurate for the
massive parameter scans it is designed to serve.

%%%%%%%%%%%%%%%%%%%%%%%%%%%%%%%%%%%%%%%%%%%%%%%%%%%%%%%%%%%%%%%%%
\section*{Acknowledgments}
{We thank the anonymous referee for the valuable and constructive suggestions.}
This work has been supported by the National Key Research and Development Program of China (No. 2023YFC2206704).
ZCC is supported by the National Natural Science Foundation of China under Grant No.~12405056, the Natural Science Foundation of Hunan Province under Grant No.~2025JJ40006, and the Innovative Research Group of Hunan Province under Grant No.~2024JJ1006.
LL is supported by the National Natural Science Foundation of China (Grant No.~12505054, 12447101 and 12433001) and the Fundamental Research Funds for the Central Universities.

\bibliographystyle{JHEP}
\bibliography{refs}

@article{Grunthal:2024sor,
    author = "Grunthal, Kathrin and others",
    title = "{The MeerKAT Pulsar Timing Array: Maps of the gravitational wave sky with the 4.5-yr data release}",
    eprint = "2412.01214",
    archivePrefix = "arXiv",
    primaryClass = "astro-ph.HE",
    doi = "10.1093/mnras/stae2573",
    journal = "Mon. Not. Roy. Astron. Soc.",
    volume = "536",
    number = "2",
    pages = "1501--1517",
    year = "2024"
}

@article{Tomita:1967wkp,
    author = "Tomita, Kenji",
    title = "{Non-Linear Theory of Gravitational Instability in the Expanding Universe}",
    doi = "10.1143/PTP.37.831",
    journal = "Prog. Theor. Phys.",
    volume = "37",
    number = "5",
    pages = "831--846",
    year = "1967"
}

@article{Matarrese:1992rp,
    author = "Matarrese, Sabino and Pantano, Ornella and Saez, Diego",
    title = "{A General relativistic approach to the nonlinear evolution of collisionless matter}",
    reportNumber = "DFPD-92-A-39",
    doi = "10.1103/PhysRevD.47.1311",
    journal = "Phys. Rev. D",
    volume = "47",
    pages = "1311--1323",
    year = "1993"
}

@article{Matarrese:1993zf,
    author = "Matarrese, Sabino and Pantano, Ornella and Saez, Diego",
    title = "{General relativistic dynamics of irrotational dust: Cosmological implications}",
    eprint = "astro-ph/9310036",
    archivePrefix = "arXiv",
    reportNumber = "DFPD-93-A-67",
    doi = "10.1103/PhysRevLett.72.320",
    journal = "Phys. Rev. Lett.",
    volume = "72",
    pages = "320--323",
    year = "1994"
}

@article{Matarrese:1997ay,
    author = "Matarrese, Sabino and Mollerach, Silvia and Bruni, Marco",
    title = "{Second order perturbations of the Einstein-de Sitter universe}",
    eprint = "astro-ph/9707278",
    archivePrefix = "arXiv",
    reportNumber = "SISSA-83-97-A",
    doi = "10.1103/PhysRevD.58.043504",
    journal = "Phys. Rev. D",
    volume = "58",
    pages = "043504",
    year = "1998"
}

@article{Mollerach:2003nq,
    author = "Mollerach, Silvia and Harari, Diego and Matarrese, Sabino",
    title = "{CMB polarization from secondary vector and tensor modes}",
    eprint = "astro-ph/0310711",
    archivePrefix = "arXiv",
    doi = "10.1103/PhysRevD.69.063002",
    journal = "Phys. Rev. D",
    volume = "69",
    pages = "063002",
    year = "2004"
}

@article{Carbone:2004iv,
    author = "Carbone, Carmelita and Matarrese, Sabino",
    title = "{A Unified treatment of cosmological perturbations from super-horizon to small scales}",
    eprint = "astro-ph/0407611",
    archivePrefix = "arXiv",
    reportNumber = "DFPD-04-A-18",
    doi = "10.1103/PhysRevD.71.043508",
    journal = "Phys. Rev. D",
    volume = "71",
    pages = "043508",
    year = "2005"
}

@article{Ananda:2006af,
    author = "Ananda, Kishore N. and Clarkson, Chris and Wands, David",
    title = "{The Cosmological gravitational wave background from primordial density perturbations}",
    eprint = "gr-qc/0612013",
    archivePrefix = "arXiv",
    doi = "10.1103/PhysRevD.75.123518",
    journal = "Phys. Rev. D",
    volume = "75",
    pages = "123518",
    year = "2007"
}

@article{Baumann:2007zm,
    author = "Baumann, Daniel and Steinhardt, Paul J. and Takahashi, Keitaro and Ichiki, Kiyotomo",
    title = "{Gravitational Wave Spectrum Induced by Primordial Scalar Perturbations}",
    eprint = "hep-th/0703290",
    archivePrefix = "arXiv",
    doi = "10.1103/PhysRevD.76.084019",
    journal = "Phys. Rev. D",
    volume = "76",
    pages = "084019",
    year = "2007"
}

@article{Saito:2008jc,
    author = "Saito, Ryo and Yokoyama, Jun'ichi",
    title = "{Gravitational wave background as a probe of the primordial black hole abundance}",
    eprint = "0812.4339",
    archivePrefix = "arXiv",
    primaryClass = "astro-ph",
    reportNumber = "RESCEU-63-08",
    doi = "10.1103/PhysRevLett.102.161101",
    journal = "Phys. Rev. Lett.",
    volume = "102",
    pages = "161101",
    year = "2009",
    note = "[Erratum: Phys.Rev.Lett. 107, 069901 (2011)]"
}

@article{Kohri:2018awv,
    author = "Kohri, Kazunori and Terada, Takahiro",
    title = "{Semianalytic calculation of gravitational wave spectrum nonlinearly induced from primordial curvature perturbations}",
    eprint = "1804.08577",
    archivePrefix = "arXiv",
    primaryClass = "gr-qc",
    reportNumber = "KEK-TH-2046, KEK-COSMO-223",
    doi = "10.1103/PhysRevD.97.123532",
    journal = "Phys. Rev. D",
    volume = "97",
    number = "12",
    pages = "123532",
    year = "2018"
}

@article{Espinosa:2018eve,
    author = "Espinosa, Jos{\'e} Ram{\'o}n and Racco, Davide and Riotto, Antonio",
    title = "{A Cosmological Signature of the SM Higgs Instability: Gravitational Waves}",
    eprint = "1804.07732",
    archivePrefix = "arXiv",
    primaryClass = "hep-ph",
    doi = "10.1088/1475-7516/2018/09/012",
    journal = "JCAP",
    volume = "09",
    pages = "012",
    year = "2018"
}

@article{Domenech:2021ztg,
    author = "Dom{\`e}nech, Guillem",
    title = "{Scalar Induced Gravitational Waves Review}",
    eprint = "2109.01398",
    archivePrefix = "arXiv",
    primaryClass = "gr-qc",
    doi = "10.3390/universe7110398",
    journal = "Universe",
    volume = "7",
    number = "11",
    pages = "398",
    year = "2021"
}

@article{Assadullahi:2009nf,
    author = "Assadullahi, Hooshyar and Wands, David",
    title = "{Gravitational waves from an early matter era}",
    eprint = "0901.0989",
    archivePrefix = "arXiv",
    primaryClass = "astro-ph.CO",
    doi = "10.1103/PhysRevD.79.083511",
    journal = "Phys. Rev. D",
    volume = "79",
    pages = "083511",
    year = "2009"
}

@article{Bugaev:2009zh,
    author = "Bugaev, Edgar and Klimai, Peter",
    title = "{Induced gravitational wave background and primordial black holes}",
    eprint = "0908.0664",
    archivePrefix = "arXiv",
    primaryClass = "astro-ph.CO",
    doi = "10.1103/PhysRevD.81.023517",
    journal = "Phys. Rev. D",
    volume = "81",
    pages = "023517",
    year = "2010"
}

@article{Bugaev:2010bb,
    author = "Bugaev, Edgar and Klimai, Peter",
    title = "{Constraints on the induced gravitational wave background from primordial black holes}",
    eprint = "1012.4697",
    archivePrefix = "arXiv",
    primaryClass = "astro-ph.CO",
    doi = "10.1103/PhysRevD.83.083521",
    journal = "Phys. Rev. D",
    volume = "83",
    pages = "083521",
    year = "2011"
}

@article{Alabidi:2012ex,
    author = "Alabidi, Laila and Kohri, Kazunori and Sasaki, Misao and Sendouda, Yuuiti",
    title = "{Observable Spectra of Induced Gravitational Waves from Inflation}",
    eprint = "1203.4663",
    archivePrefix = "arXiv",
    primaryClass = "astro-ph.CO",
    doi = "10.1088/1475-7516/2012/09/017",
    journal = "JCAP",
    volume = "09",
    pages = "017",
    year = "2012"
}

@article{Alabidi:2013lya,
    author = "Alabidi, Laila and Kohri, Kazunori and Sasaki, Misao and Sendouda, Yuuiti",
    title = "{Observable induced gravitational waves from an early matter phase}",
    eprint = "1303.4519",
    archivePrefix = "arXiv",
    primaryClass = "astro-ph.CO",
    doi = "10.1088/1475-7516/2013/05/033",
    journal = "JCAP",
    volume = "05",
    pages = "033",
    year = "2013"
}

@article{Nakama:2016gzw,
    author = "Nakama, Tomohiro and Silk, Joseph and Kamionkowski, Marc",
    title = "{Stochastic gravitational waves associated with the formation of primordial black holes}",
    eprint = "1612.06264",
    archivePrefix = "arXiv",
    primaryClass = "astro-ph.CO",
    doi = "10.1103/PhysRevD.95.043511",
    journal = "Phys. Rev. D",
    volume = "95",
    number = "4",
    pages = "043511",
    year = "2017"
}

@article{Inomata:2018epa,
    author = "Inomata, Keisuke and Nakama, Tomohiro",
    title = "{Gravitational waves induced by scalar perturbations as probes of the small-scale primordial spectrum}",
    eprint = "1812.00674",
    archivePrefix = "arXiv",
    primaryClass = "astro-ph.CO",
    reportNumber = "IPMU 18-0200",
    doi = "10.1103/PhysRevD.99.043511",
    journal = "Phys. Rev. D",
    volume = "99",
    number = "4",
    pages = "043511",
    year = "2019"
}

@article{Yuan:2019udt,
    author = "Yuan, Chen and Chen, Zu-Cheng and Huang, Qing-Guo",
    title = "{Probing primordial{\textendash}black-hole dark matter with scalar induced gravitational waves}",
    eprint = "1906.11549",
    archivePrefix = "arXiv",
    primaryClass = "astro-ph.CO",
    doi = "10.1103/PhysRevD.100.081301",
    journal = "Phys. Rev. D",
    volume = "100",
    number = "8",
    pages = "8",
    year = "2019"
}

@article{Hawking:1971ei,
    author = "Hawking, Stephen",
    title = "{Gravitationally collapsed objects of very low mass}",
    doi = "10.1093/mnras/152.1.75",
    journal = "Mon. Not. Roy. Astron. Soc.",
    volume = "152",
    pages = "75",
    year = "1971"
}

@article{Carr:1974nx,
    author = "Carr, Bernard J. and Hawking, S. W.",
    title = "{Black holes in the early Universe}",
    doi = "10.1093/mnras/168.2.399",
    journal = "Mon. Not. Roy. Astron. Soc.",
    volume = "168",
    pages = "399--415",
    year = "1974"
}

@article{Carr:1975qj,
    author = "Carr, Bernard J.",
    title = "{The Primordial black hole mass spectrum}",
    doi = "10.1086/153853",
    journal = "Astrophys. J.",
    volume = "201",
    pages = "1--19",
    year = "1975"
}

@article{Ivanov:1994pa,
    author = "Ivanov, P. and Naselsky, P. and Novikov, I.",
    title = "{Inflation and primordial black holes as dark matter}",
    reportNumber = "NORDITA-94-12-A",
    doi = "10.1103/PhysRevD.50.7173",
    journal = "Phys. Rev. D",
    volume = "50",
    pages = "7173--7178",
    year = "1994"
}

@article{Garcia-Bellido:1996mdl,
    author = "Garcia-Bellido, Juan and Linde, Andrei D. and Wands, David",
    title = "{Density perturbations and black hole formation in hybrid inflation}",
    eprint = "astro-ph/9605094",
    archivePrefix = "arXiv",
    reportNumber = "SU-ITP-96-20, SUSSEX-AST-96-5-1",
    doi = "10.1103/PhysRevD.54.6040",
    journal = "Phys. Rev. D",
    volume = "54",
    pages = "6040--6058",
    year = "1996"
}

@article{Cai:2018dig,
    author = "Cai, Rong-gen and Pi, Shi and Sasaki, Misao",
    title = "{Gravitational Waves Induced by non-Gaussian Scalar Perturbations}",
    eprint = "1810.11000",
    archivePrefix = "arXiv",
    primaryClass = "astro-ph.CO",
    reportNumber = "IPMU18-0172, YITP-18-114",
    doi = "10.1103/PhysRevLett.122.201101",
    journal = "Phys. Rev. Lett.",
    volume = "122",
    number = "20",
    pages = "201101",
    year = "2019"
}

@article{Sasaki:2018dmp,
    author = "Sasaki, Misao and Suyama, Teruaki and Tanaka, Takahiro and Yokoyama, Shuichiro",
    title = "{Primordial black holes{\textemdash}perspectives in gravitational wave astronomy}",
    eprint = "1801.05235",
    archivePrefix = "arXiv",
    primaryClass = "astro-ph.CO",
    doi = "10.1088/1361-6382/aaa7b4",
    journal = "Class. Quant. Grav.",
    volume = "35",
    number = "6",
    pages = "063001",
    year = "2018"
}

@article{Sasaki:2016jop,
    author = "Sasaki, Misao and Suyama, Teruaki and Tanaka, Takahiro and Yokoyama, Shuichiro",
    title = "{Primordial Black Hole Scenario for the Gravitational-Wave Event GW150914}",
    eprint = "1603.08338",
    archivePrefix = "arXiv",
    primaryClass = "astro-ph.CO",
    reportNumber = "RESCEU-17-16, RUP-16-7, YITP-16-43",
    doi = "10.1103/PhysRevLett.117.061101",
    journal = "Phys. Rev. Lett.",
    volume = "117",
    number = "6",
    pages = "061101",
    year = "2016",
    note = "[Erratum: Phys.Rev.Lett. 121, 059901 (2018)]"
}

@article{Bird:2016dcv,
    author = {Bird, Simeon and Cholis, Ilias and Mu{\~n}oz, Julian B. and Ali-Ha{\"\i}moud, Yacine and Kamionkowski, Marc and Kovetz, Ely D. and Raccanelli, Alvise and Riess, Adam G.},
    title = "{Did LIGO detect dark matter?}",
    eprint = "1603.00464",
    archivePrefix = "arXiv",
    primaryClass = "astro-ph.CO",
    doi = "10.1103/PhysRevLett.116.201301",
    journal = "Phys. Rev. Lett.",
    volume = "116",
    number = "20",
    pages = "201301",
    year = "2016"
}

@article{Clesse:2016vqa,
    author = "Clesse, Sebastien and Garc{\'\i}a-Bellido, Juan",
    title = "{The clustering of massive Primordial Black Holes as Dark Matter: measuring their mass distribution with Advanced LIGO}",
    eprint = "1603.05234",
    archivePrefix = "arXiv",
    primaryClass = "astro-ph.CO",
    reportNumber = "TTK-16-10, IFT-UAM-CSIC-16-027",
    doi = "10.1016/j.dark.2016.10.002",
    journal = "Phys. Dark Univ.",
    volume = "15",
    pages = "142--147",
    year = "2017"
}

@article{Carr:2016drx,
    author = "Carr, Bernard and Kuhnel, Florian and Sandstad, Marit",
    title = "{Primordial Black Holes as Dark Matter}",
    eprint = "1607.06077",
    archivePrefix = "arXiv",
    primaryClass = "astro-ph.CO",
    reportNumber = "NORDITA-2016-83",
    doi = "10.1103/PhysRevD.94.083504",
    journal = "Phys. Rev. D",
    volume = "94",
    number = "8",
    pages = "083504",
    year = "2016"
}

@article{Ali-Haimoud:2017rtz,
    author = {Ali-Ha{\"\i}moud, Yacine and Kovetz, Ely D. and Kamionkowski, Marc},
    title = "{Merger rate of primordial black-hole binaries}",
    eprint = "1709.06576",
    archivePrefix = "arXiv",
    primaryClass = "astro-ph.CO",
    doi = "10.1103/PhysRevD.96.123523",
    journal = "Phys. Rev. D",
    volume = "96",
    number = "12",
    pages = "123523",
    year = "2017"
}

@article{Raidal:2017mfl,
    author = {Raidal, Martti and Vaskonen, Ville and Veerm{\"a}e, Hardi},
    title = "{Gravitational Waves from Primordial Black Hole Mergers}",
    eprint = "1707.01480",
    archivePrefix = "arXiv",
    primaryClass = "astro-ph.CO",
    doi = "10.1088/1475-7516/2017/09/037",
    journal = "JCAP",
    volume = "09",
    pages = "037",
    year = "2017"
}

@article{Chen:2018czv,
    author = "Chen, Zu-Cheng and Huang, Qing-Guo",
    title = "{Merger Rate Distribution of Primordial-Black-Hole Binaries}",
    eprint = "1801.10327",
    archivePrefix = "arXiv",
    primaryClass = "astro-ph.CO",
    doi = "10.3847/1538-4357/aad6e2",
    journal = "Astrophys. J.",
    volume = "864",
    number = "1",
    pages = "61",
    year = "2018"
}

@article{Liu:2018ess,
    author = "Liu, Lang and Guo, Zong-Kuan and Cai, Rong-Gen",
    title = "{Effects of the surrounding primordial black holes on the merger rate of primordial black hole binaries}",
    eprint = "1812.05376",
    archivePrefix = "arXiv",
    primaryClass = "astro-ph.CO",
    doi = "10.1103/PhysRevD.99.063523",
    journal = "Phys. Rev. D",
    volume = "99",
    number = "6",
    pages = "063523",
    year = "2019"
}

@article{Vaskonen:2019jpv,
    author = {Vaskonen, Ville and Veerm{\"a}e, Hardi},
    title = "{Lower bound on the primordial black hole merger rate}",
    eprint = "1908.09752",
    archivePrefix = "arXiv",
    primaryClass = "astro-ph.CO",
    reportNumber = "CERN-TH-2019-141, KCL-PH-TH/2019-69",
    doi = "10.1103/PhysRevD.101.043015",
    journal = "Phys. Rev. D",
    volume = "101",
    number = "4",
    pages = "043015",
    year = "2020"
}

@article{DeLuca:2020qqa,
    author = "De Luca, V. and Franciolini, G. and Pani, P. and Riotto, A.",
    title = "{Primordial Black Holes Confront LIGO/Virgo data: Current situation}",
    eprint = "2005.05641",
    archivePrefix = "arXiv",
    primaryClass = "astro-ph.CO",
    doi = "10.1088/1475-7516/2020/06/044",
    journal = "JCAP",
    volume = "06",
    pages = "044",
    year = "2020"
}

@article{Hutsi:2020sol,
    author = {H{\"u}tsi, Gert and Raidal, Martti and Vaskonen, Ville and Veerm{\"a}e, Hardi},
    title = "{Two populations of LIGO-Virgo black holes}",
    eprint = "2012.02786",
    archivePrefix = "arXiv",
    primaryClass = "astro-ph.CO",
    doi = "10.1088/1475-7516/2021/03/068",
    journal = "JCAP",
    volume = "03",
    pages = "068",
    year = "2021"
}

@article{Franciolini:2021tla,
    author = "Franciolini, Gabriele and Baibhav, Vishal and De Luca, Valerio and Ng, Ken K. Y. and Wong, Kaze W. K. and Berti, Emanuele and Pani, Paolo and Riotto, Antonio and Vitale, Salvatore",
    title = "{Searching for a subpopulation of primordial black holes in LIGO-Virgo gravitational-wave data}",
    eprint = "2105.03349",
    archivePrefix = "arXiv",
    primaryClass = "gr-qc",
    doi = "10.1103/PhysRevD.105.083526",
    journal = "Phys. Rev. D",
    volume = "105",
    number = "8",
    pages = "083526",
    year = "2022"
}

@article{Carr:2020gox,
    author = "Carr, Bernard and Kohri, Kazunori and Sendouda, Yuuiti and Yokoyama, Jun'ichi",
    title = "{Constraints on primordial black holes}",
    eprint = "2002.12778",
    archivePrefix = "arXiv",
    primaryClass = "astro-ph.CO",
    reportNumber = "RESCEU-03/20; KEK-Cosmo-249; KEK-TH-2199; IPMU20-0024",
    doi = "10.1088/1361-6633/ac1e31",
    journal = "Rept. Prog. Phys.",
    volume = "84",
    number = "11",
    pages = "116902",
    year = "2021"
}

@article{Green:2020jor,
    author = "Green, Anne M. and Kavanagh, Bradley J.",
    title = "{Primordial Black Holes as a dark matter candidate}",
    eprint = "2007.10722",
    archivePrefix = "arXiv",
    primaryClass = "astro-ph.CO",
    doi = "10.1088/1361-6471/abc534",
    journal = "J. Phys. G",
    volume = "48",
    number = "4",
    pages = "043001",
    year = "2021"
}

@article{Escriva:2022duf,
    author = "Escriv{\`a}, Albert and Kuhnel, Florian and Tada, Yuichiro",
    editor = "Sedda, Manuel Arca and Bortolas, Elisa and Spera, Mario",
    title = "{Primordial Black Holes}",
    eprint = "2211.05767",
    archivePrefix = "arXiv",
    primaryClass = "astro-ph.CO",
    doi = "10.1016/B978-0-32-395636-9.00012-8",
    month = "11",
    year = "2022"
}

@article{Carr:2020xqk,
    author = "Carr, Bernard and Kuhnel, Florian",
    title = "{Primordial Black Holes as Dark Matter: Recent Developments}",
    eprint = "2006.02838",
    archivePrefix = "arXiv",
    primaryClass = "astro-ph.CO",
    doi = "10.1146/annurev-nucl-050520-125911",
    journal = "Ann. Rev. Nucl. Part. Sci.",
    volume = "70",
    pages = "355--394",
    year = "2020"
}

@article{Villanueva-Domingo:2021spv,
    author = "Villanueva-Domingo, Pablo and Mena, Olga and Palomares-Ruiz, Sergio",
    title = "{A brief review on primordial black holes as dark matter}",
    eprint = "2103.12087",
    archivePrefix = "arXiv",
    primaryClass = "astro-ph.CO",
    doi = "10.3389/fspas.2021.681084",
    journal = "Front. Astron. Space Sci.",
    volume = "8",
    pages = "87",
    year = "2021"
}

@article{Young:2014ana,
    author = "Young, Sam and Byrnes, Christian T. and Sasaki, Misao",
    title = "{Calculating the mass fraction of primordial black holes}",
    eprint = "1405.7023",
    archivePrefix = "arXiv",
    primaryClass = "gr-qc",
    doi = "10.1088/1475-7516/2014/07/045",
    journal = "JCAP",
    volume = "07",
    pages = "045",
    year = "2014"
}

@article{Germani:2018jgr,
    author = "Germani, Cristiano and Musco, Ilia",
    title = "{Abundance of Primordial Black Holes Depends on the Shape of the Inflationary Power Spectrum}",
    eprint = "1805.04087",
    archivePrefix = "arXiv",
    primaryClass = "astro-ph.CO",
    reportNumber = "ICCUB-18-009",
    doi = "10.1103/PhysRevLett.122.141302",
    journal = "Phys. Rev. Lett.",
    volume = "122",
    number = "14",
    pages = "141302",
    year = "2019"
}

@article{Musco:2018rwt,
    author = "Musco, Ilia",
    title = "{Threshold for primordial black holes: Dependence on the shape of the cosmological perturbations}",
    eprint = "1809.02127",
    archivePrefix = "arXiv",
    primaryClass = "gr-qc",
    doi = "10.1103/PhysRevD.100.123524",
    journal = "Phys. Rev. D",
    volume = "100",
    number = "12",
    pages = "123524",
    year = "2019"
}

@article{Escriva:2019phb,
    author = "Escriv{\`a}, Albert and Germani, Cristiano and Sheth, Ravi K.",
    title = "{Universal threshold for primordial black hole formation}",
    eprint = "1907.13311",
    archivePrefix = "arXiv",
    primaryClass = "gr-qc",
    reportNumber = "ICC-19-013",
    doi = "10.1103/PhysRevD.101.044022",
    journal = "Phys. Rev. D",
    volume = "101",
    number = "4",
    pages = "044022",
    year = "2020"
}

@article{Carr:2009jm,
    author = "Carr, B. J. and Kohri, Kazunori and Sendouda, Yuuiti and Yokoyama, Jun'ichi",
    title = "{New cosmological constraints on primordial black holes}",
    eprint = "0912.5297",
    archivePrefix = "arXiv",
    primaryClass = "astro-ph.CO",
    reportNumber = "RESCEU-31-09, TU-852, YITP-09-112",
    doi = "10.1103/PhysRevD.81.104019",
    journal = "Phys. Rev. D",
    volume = "81",
    pages = "104019",
    year = "2010"
}

@article{Niikura:2017zjd,
    author = "Niikura, Hiroko and others",
    title = "{Microlensing constraints on primordial black holes with Subaru/HSC Andromeda observations}",
    eprint = "1701.02151",
    archivePrefix = "arXiv",
    primaryClass = "astro-ph.CO",
    doi = "10.1038/s41550-019-0723-1",
    journal = "Nature Astron.",
    volume = "3",
    number = "6",
    pages = "524--534",
    year = "2019"
}

@article{Garcia-Bellido:2017mdw,
    author = "Garcia-Bellido, Juan and Ruiz Morales, Ester",
    title = "{Primordial black holes from single field models of inflation}",
    eprint = "1702.03901",
    archivePrefix = "arXiv",
    primaryClass = "astro-ph.CO",
    reportNumber = "IFT-UAM-CSIC-17-007, CERN-TH-2017-196",
    doi = "10.1016/j.dark.2017.09.007",
    journal = "Phys. Dark Univ.",
    volume = "18",
    pages = "47--54",
    year = "2017"
}

@article{Pi:2017gih,
    author = "Pi, Shi and Zhang, Ying-li and Huang, Qing-Guo and Sasaki, Misao",
    title = "{Scalaron from $R^2$-gravity as a heavy field}",
    eprint = "1712.09896",
    archivePrefix = "arXiv",
    primaryClass = "astro-ph.CO",
    reportNumber = "YITP-17-135",
    doi = "10.1088/1475-7516/2018/05/042",
    journal = "JCAP",
    volume = "05",
    pages = "042",
    year = "2018"
}

@article{Inomata:2017okj,
    author = "Inomata, Keisuke and Kawasaki, Masahiro and Mukaida, Kyohei and Tada, Yuichiro and Yanagida, Tsutomu T.",
    title = "{Inflationary Primordial Black Holes as All Dark Matter}",
    eprint = "1701.02544",
    archivePrefix = "arXiv",
    primaryClass = "astro-ph.CO",
    reportNumber = "IPMU17-0009",
    doi = "10.1103/PhysRevD.96.043504",
    journal = "Phys. Rev. D",
    volume = "96",
    number = "4",
    pages = "043504",
    year = "2017"
}

@article{Espinosa:2017sgp,
    author = "Espinosa, J. R. and Racco, D. and Riotto, A.",
    title = "{Cosmological Signature of the Standard Model Higgs Vacuum Instability: Primordial Black Holes as Dark Matter}",
    eprint = "1710.11196",
    archivePrefix = "arXiv",
    primaryClass = "hep-ph",
    doi = "10.1103/PhysRevLett.120.121301",
    journal = "Phys. Rev. Lett.",
    volume = "120",
    number = "12",
    pages = "121301",
    year = "2018"
}

@article{Kannike:2017bxn,
    author = {Kannike, Kristjan and Marzola, Luca and Raidal, Martti and Veerm{\"a}e, Hardi},
    title = "{Single Field Double Inflation and Primordial Black Holes}",
    eprint = "1705.06225",
    archivePrefix = "arXiv",
    primaryClass = "astro-ph.CO",
    doi = "10.1088/1475-7516/2017/09/020",
    journal = "JCAP",
    volume = "09",
    pages = "020",
    year = "2017"
}

@article{Cai:2018tuh,
    author = "Cai, Yi-Fu and Tong, Xi and Wang, Dong-Gang and Yan, Sheng-Feng",
    title = "{Primordial Black Holes from Sound Speed Resonance during Inflation}",
    eprint = "1805.03639",
    archivePrefix = "arXiv",
    primaryClass = "astro-ph.CO",
    doi = "10.1103/PhysRevLett.121.081306",
    journal = "Phys. Rev. Lett.",
    volume = "121",
    number = "8",
    pages = "081306",
    year = "2018"
}

@article{Byrnes:2018txb,
    author = "Byrnes, Christian T. and Cole, Philippa S. and Patil, Subodh P.",
    title = "{Steepest growth of the power spectrum and primordial black holes}",
    eprint = "1811.11158",
    archivePrefix = "arXiv",
    primaryClass = "astro-ph.CO",
    doi = "10.1088/1475-7516/2019/06/028",
    journal = "JCAP",
    volume = "06",
    pages = "028",
    year = "2019"
}

@article{Motohashi:2017kbs,
    author = "Motohashi, Hayato and Hu, Wayne",
    title = "{Primordial Black Holes and Slow-Roll Violation}",
    eprint = "1706.06784",
    archivePrefix = "arXiv",
    primaryClass = "astro-ph.CO",
    doi = "10.1103/PhysRevD.96.063503",
    journal = "Phys. Rev. D",
    volume = "96",
    number = "6",
    pages = "063503",
    year = "2017"
}

@article{Ezquiaga:2017fvi,
    author = "Ezquiaga, Jose Maria and Garcia-Bellido, Juan and Ruiz Morales, Ester",
    title = "{Primordial Black Hole production in Critical Higgs Inflation}",
    eprint = "1705.04861",
    archivePrefix = "arXiv",
    primaryClass = "astro-ph.CO",
    reportNumber = "IFT-UAM-CSIC-17-043",
    doi = "10.1016/j.physletb.2017.11.039",
    journal = "Phys. Lett. B",
    volume = "776",
    pages = "345--349",
    year = "2018"
}

@article{Di:2017ndc,
    author = "Di, Haoran and Gong, Yungui",
    title = "{Primordial black holes and second order gravitational waves from ultra-slow-roll inflation}",
    eprint = "1707.09578",
    archivePrefix = "arXiv",
    primaryClass = "astro-ph.CO",
    doi = "10.1088/1475-7516/2018/07/007",
    journal = "JCAP",
    volume = "07",
    pages = "007",
    year = "2018"
}

@article{Dalianis:2018frf,
    author = "Dalianis, Ioannis and Kehagias, Alex and Tringas, George",
    title = "{Primordial black holes from {\ensuremath{\alpha}}-attractors}",
    eprint = "1805.09483",
    archivePrefix = "arXiv",
    primaryClass = "astro-ph.CO",
    doi = "10.1088/1475-7516/2019/01/037",
    journal = "JCAP",
    volume = "01",
    pages = "037",
    year = "2019"
}

@article{Bhaumik:2019tvl,
    author = "Bhaumik, Nilanjandev and Jain, Rajeev Kumar",
    title = "{Primordial black holes dark matter from inflection point models of inflation and the effects of reheating}",
    eprint = "1907.04125",
    archivePrefix = "arXiv",
    primaryClass = "astro-ph.CO",
    doi = "10.1088/1475-7516/2020/01/037",
    journal = "JCAP",
    volume = "01",
    pages = "037",
    year = "2020"
}

@article{Mishra:2019pzq,
    author = "Mishra, Swagat S. and Sahni, Varun",
    title = "{Primordial Black Holes from a tiny bump/dip in the Inflaton potential}",
    eprint = "1911.00057",
    archivePrefix = "arXiv",
    primaryClass = "gr-qc",
    doi = "10.1088/1475-7516/2020/04/007",
    journal = "JCAP",
    volume = "04",
    pages = "007",
    year = "2020"
}

@article{Liu:2020oqe,
    author = "Liu, Jing and Guo, Zong-Kuan and Cai, Rong-Gen",
    title = "{Analytical approximation of the scalar spectrum in the ultraslow-roll inflationary models}",
    eprint = "2003.02075",
    archivePrefix = "arXiv",
    primaryClass = "astro-ph.CO",
    doi = "10.1103/PhysRevD.101.083535",
    journal = "Phys. Rev. D",
    volume = "101",
    number = "8",
    pages = "083535",
    year = "2020"
}

@article{Fu:2019ttf,
    author = "Fu, Chengjie and Wu, Puxun and Yu, Hongwei",
    title = "{Primordial Black Holes from Inflation with Nonminimal Derivative Coupling}",
    eprint = "1907.05042",
    archivePrefix = "arXiv",
    primaryClass = "astro-ph.CO",
    doi = "10.1103/PhysRevD.100.063532",
    journal = "Phys. Rev. D",
    volume = "100",
    number = "6",
    pages = "063532",
    year = "2019"
}

@article{Fu:2020lob,
    author = "Fu, Chengjie and Wu, Puxun and Yu, Hongwei",
    title = "{Primordial black holes and oscillating gravitational waves in slow-roll and slow-climb inflation with an intermediate noninflationary phase}",
    eprint = "2006.03768",
    archivePrefix = "arXiv",
    primaryClass = "astro-ph.CO",
    doi = "10.1103/PhysRevD.102.043527",
    journal = "Phys. Rev. D",
    volume = "102",
    number = "4",
    pages = "043527",
    year = "2020"
}

@article{Inomata:2021uqj,
    author = "Inomata, Keisuke and McDonough, Evan and Hu, Wayne",
    title = "{Primordial black holes arise when the inflaton falls}",
    eprint = "2104.03972",
    archivePrefix = "arXiv",
    primaryClass = "astro-ph.CO",
    doi = "10.1103/PhysRevD.104.123553",
    journal = "Phys. Rev. D",
    volume = "104",
    number = "12",
    pages = "123553",
    year = "2021"
}

@article{Kawai:2021edk,
    author = "Kawai, Shinsuke and Kim, Jinsu",
    title = "{Primordial black holes from Gauss-Bonnet-corrected single field inflation}",
    eprint = "2108.01340",
    archivePrefix = "arXiv",
    primaryClass = "astro-ph.CO",
    reportNumber = "CERN-TH-2021-115",
    doi = "10.1103/PhysRevD.104.083545",
    journal = "Phys. Rev. D",
    volume = "104",
    number = "8",
    pages = "083545",
    year = "2021"
}

@article{Ballesteros:2020qam,
    author = "Ballesteros, Guillermo and Rey, Juli{\'a}n and Taoso, Marco and Urbano, Alfredo",
    title = "{Primordial black holes as dark matter and gravitational waves from single-field polynomial inflation}",
    eprint = "2001.08220",
    archivePrefix = "arXiv",
    primaryClass = "astro-ph.CO",
    doi = "10.1088/1475-7516/2020/07/025",
    journal = "JCAP",
    volume = "07",
    pages = "025",
    year = "2020"
}

@article{Ragavendra:2020sop,
    author = "Ragavendra, H. V. and Saha, Pankaj and Sriramkumar, L. and Silk, Joseph",
    title = "{Primordial black holes and secondary gravitational waves from ultraslow roll and punctuated inflation}",
    eprint = "2008.12202",
    archivePrefix = "arXiv",
    primaryClass = "astro-ph.CO",
    doi = "10.1103/PhysRevD.103.083510",
    journal = "Phys. Rev. D",
    volume = "103",
    number = "8",
    pages = "083510",
    year = "2021"
}

@article{Ozsoy:2018flq,
    author = {{\"O}zsoy, Ogan and Parameswaran, Susha and Tasinato, Gianmassimo and Zavala, Ivonne},
    title = "{Mechanisms for Primordial Black Hole Production in String Theory}",
    eprint = "1803.07626",
    archivePrefix = "arXiv",
    primaryClass = "hep-th",
    doi = "10.1088/1475-7516/2018/07/005",
    journal = "JCAP",
    volume = "07",
    pages = "005",
    year = "2018"
}

@article{Cicoli:2018asa,
    author = "Cicoli, Michele and Diaz, Victor A. and Pedro, Francisco G.",
    title = "{Primordial Black Holes from String Inflation}",
    eprint = "1803.02837",
    archivePrefix = "arXiv",
    primaryClass = "hep-th",
    doi = "10.1088/1475-7516/2018/06/034",
    journal = "JCAP",
    volume = "06",
    pages = "034",
    year = "2018"
}

@article{Ozsoy:2023ryl,
    author = {{\"O}zsoy, Ogan and Tasinato, Gianmassimo},
    title = "{Inflation and Primordial Black Holes}",
    eprint = "2301.03600",
    archivePrefix = "arXiv",
    primaryClass = "astro-ph.CO",
    doi = "10.3390/universe9050203",
    journal = "Universe",
    volume = "9",
    number = "5",
    pages = "203",
    year = "2023"
}

@article{Pattison:2017mbe,
    author = "Pattison, Chris and Vennin, Vincent and Assadullahi, Hooshyar and Wands, David",
    title = "{Quantum diffusion during inflation and primordial black holes}",
    eprint = "1707.00537",
    archivePrefix = "arXiv",
    primaryClass = "hep-th",
    doi = "10.1088/1475-7516/2017/10/046",
    journal = "JCAP",
    volume = "10",
    pages = "046",
    year = "2017"
}

@article{Biagetti:2018pjj,
    author = "Biagetti, Matteo and Franciolini, Gabriele and Kehagias, Alex and Riotto, Antonio",
    title = "{Primordial Black Holes from Inflation and Quantum Diffusion}",
    eprint = "1804.07124",
    archivePrefix = "arXiv",
    primaryClass = "astro-ph.CO",
    doi = "10.1088/1475-7516/2018/07/032",
    journal = "JCAP",
    volume = "07",
    pages = "032",
    year = "2018"
}

@article{Ezquiaga:2019ftu,
    author = "Ezquiaga, Jose Mar{\'\i}a and Garc{\'\i}a-Bellido, Juan and Vennin, Vincent",
    title = "{The exponential tail of inflationary fluctuations: consequences for primordial black holes}",
    eprint = "1912.05399",
    archivePrefix = "arXiv",
    primaryClass = "astro-ph.CO",
    doi = "10.1088/1475-7516/2020/03/029",
    journal = "JCAP",
    volume = "03",
    pages = "029",
    year = "2020"
}

@article{Cole:2017gle,
    author = "Cole, Philippa S. and Byrnes, Christian T.",
    title = "{Extreme scenarios: the tightest possible constraints on the power spectrum due to primordial black holes}",
    eprint = "1706.10288",
    archivePrefix = "arXiv",
    primaryClass = "astro-ph.CO",
    doi = "10.1088/1475-7516/2018/02/019",
    journal = "JCAP",
    volume = "02",
    pages = "019",
    year = "2018"
}

@article{LISA:2017pwj,
    author = "Amaro-Seoane, Pau and others",
    collaboration = "LISA",
    title = "{Laser Interferometer Space Antenna}",
    eprint = "1702.00786",
    archivePrefix = "arXiv",
    primaryClass = "astro-ph.IM",
    month = "2",
    year = "2017"
}

@article{Kawamura:2011zz,
    author = "Kawamura, Seiji and others",
    editor = "Buchman, Sasha and Sun, Ke-Xun",
    title = "{The Japanese space gravitational wave antenna: DECIGO}",
    doi = "10.1088/0264-9381/28/9/094011",
    journal = "Class. Quant. Grav.",
    volume = "28",
    pages = "094011",
    year = "2011"
}

@article{Kawamura:2020pcg,
    author = "Kawamura, Seiji and others",
    title = "{Current status of space gravitational wave antenna DECIGO and B-DECIGO}",
    eprint = "2006.13545",
    archivePrefix = "arXiv",
    primaryClass = "gr-qc",
    doi = "10.1093/ptep/ptab019",
    journal = "PTEP",
    volume = "2021",
    number = "5",
    pages = "05A105",
    year = "2021"
}

@article{Punturo:2010zz,
    author = "Punturo, M. and others",
    editor = "Ricci, Fulvio",
    title = "{The Einstein Telescope: A third-generation gravitational wave observatory}",
    doi = "10.1088/0264-9381/27/19/194002",
    journal = "Class. Quant. Grav.",
    volume = "27",
    pages = "194002",
    year = "2010"
}

@article{Maggiore:2019uih,
    author = "Maggiore, Michele and others",
    collaboration = "ET",
    title = "{Science Case for the Einstein Telescope}",
    eprint = "1912.02622",
    archivePrefix = "arXiv",
    primaryClass = "astro-ph.CO",
    doi = "10.1088/1475-7516/2020/03/050",
    journal = "JCAP",
    volume = "03",
    pages = "050",
    year = "2020"
}

@article{Reitze:2019iox,
    author = "Reitze, David and others",
    title = "{Cosmic Explorer: The U.S. Contribution to Gravitational-Wave Astronomy beyond LIGO}",
    eprint = "1907.04833",
    archivePrefix = "arXiv",
    primaryClass = "astro-ph.IM",
    reportNumber = "LIGO-P1900316",
    journal = "Bull. Am. Astron. Soc.",
    volume = "51",
    number = "7",
    pages = "035",
    year = "2019"
}

@article{Crowder:2005nr,
    author = "Crowder, Jeff and Cornish, Neil J.",
    title = "{Beyond LISA: Exploring future gravitational wave missions}",
    eprint = "gr-qc/0506015",
    archivePrefix = "arXiv",
    doi = "10.1103/PhysRevD.72.083005",
    journal = "Phys. Rev. D",
    volume = "72",
    pages = "083005",
    year = "2005"
}

@article{Ruan:2018tsw,
    author = "Ruan, Wen-Hong and Guo, Zong-Kuan and Cai, Rong-Gen and Zhang, Yuan-Zhong",
    title = "{Taiji program: Gravitational-wave sources}",
    eprint = "1807.09495",
    archivePrefix = "arXiv",
    primaryClass = "gr-qc",
    doi = "10.1142/S0217751X2050075X",
    journal = "Int. J. Mod. Phys. A",
    volume = "35",
    number = "17",
    pages = "2050075",
    year = "2020"
}

@article{TianQin:2015yph,
    author = "Luo, Jun and others",
    collaboration = "TianQin",
    title = "{TianQin: a space-borne gravitational wave detector}",
    eprint = "1512.02076",
    archivePrefix = "arXiv",
    primaryClass = "astro-ph.IM",
    doi = "10.1088/0264-9381/33/3/035010",
    journal = "Class. Quant. Grav.",
    volume = "33",
    number = "3",
    pages = "035010",
    year = "2016"
}

@article{Janssen:2014dka,
    author = "Janssen, Gemma and others",
    editor = "Bourke, Tyler L. and others",
    title = "{Gravitational wave astronomy with the SKA}",
    eprint = "1501.00127",
    archivePrefix = "arXiv",
    primaryClass = "astro-ph.IM",
    doi = "10.22323/1.215.0037",
    journal = "PoS",
    volume = "AASKA14",
    pages = "037",
    year = "2015"
}

@article{Hobbs:2009yy,
    author = "Hobbs, G. and others",
    editor = "Marka, Zsuzsa and Marka, Szabolcs",
    title = "{The international pulsar timing array project: using pulsars as a gravitational wave detector}",
    eprint = "0911.5206",
    archivePrefix = "arXiv",
    primaryClass = "astro-ph.SR",
    doi = "10.1088/0264-9381/27/8/084013",
    journal = "Class. Quant. Grav.",
    volume = "27",
    pages = "084013",
    year = "2010"
}

@article{NANOGrav:2023gor,
    author = "Agazie, Gabriella and others",
    collaboration = "NANOGrav",
    title = "{The NANOGrav 15 yr Data Set: Evidence for a Gravitational-wave Background}",
    eprint = "2306.16213",
    archivePrefix = "arXiv",
    primaryClass = "astro-ph.HE",
    doi = "10.3847/2041-8213/acdac6",
    journal = "Astrophys. J. Lett.",
    volume = "951",
    number = "1",
    pages = "L8",
    year = "2023"
}

@article{EPTA:2023fyk,
    author = "Antoniadis, J. and others",
    collaboration = "EPTA, InPTA:",
    title = "{The second data release from the European Pulsar Timing Array - III. Search for gravitational wave signals}",
    eprint = "2306.16214",
    archivePrefix = "arXiv",
    primaryClass = "astro-ph.HE",
    doi = "10.1051/0004-6361/202346844",
    journal = "Astron. Astrophys.",
    volume = "678",
    pages = "A50",
    year = "2023"
}

@article{Reardon:2023gzh,
    author = "Reardon, Daniel J. and others",
    title = "{Search for an Isotropic Gravitational-wave Background with the Parkes Pulsar Timing Array}",
    eprint = "2306.16215",
    archivePrefix = "arXiv",
    primaryClass = "astro-ph.HE",
    doi = "10.3847/2041-8213/acdd02",
    journal = "Astrophys. J. Lett.",
    volume = "951",
    number = "1",
    pages = "L6",
    year = "2023"
}

@article{Xu:2023wog,
    author = "Xu, Heng and others",
    title = "{Searching for the Nano-Hertz Stochastic Gravitational Wave Background with the Chinese Pulsar Timing Array Data Release I}",
    eprint = "2306.16216",
    archivePrefix = "arXiv",
    primaryClass = "astro-ph.HE",
    doi = "10.1088/1674-4527/acdfa5",
    journal = "Res. Astron. Astrophys.",
    volume = "23",
    number = "7",
    pages = "075024",
    year = "2023"
}

@article{NANOGrav:2023hvm,
    author = "Afzal, Adeela and others",
    collaboration = "NANOGrav",
    title = "{The NANOGrav 15 yr Data Set: Search for Signals from New Physics}",
    eprint = "2306.16219",
    archivePrefix = "arXiv",
    primaryClass = "astro-ph.HE",
    reportNumber = "FERMILAB-PUB-23-589-T",
    doi = "10.3847/2041-8213/acdc91",
    journal = "Astrophys. J. Lett.",
    volume = "951",
    number = "1",
    pages = "L11",
    year = "2023",
    note = "[Erratum: Astrophys.J.Lett. 971, L27 (2024), Erratum: Astrophys.J. 971, L27 (2024)]"
}

@article{Dandoy:2023jot,
    author = "Dandoy, Virgile and Domcke, Valerie and Rompineve, Fabrizio",
    title = "{Search for scalar induced gravitational waves in the international pulsar timing array data release 2 and NANOgrav 12.5 years datasets}",
    eprint = "2302.07901",
    archivePrefix = "arXiv",
    primaryClass = "astro-ph.CO",
    reportNumber = "CERN-TH-2023-027",
    doi = "10.21468/SciPostPhysCore.6.3.060",
    journal = "SciPost Phys. Core",
    volume = "6",
    pages = "060",
    year = "2023"
}

@article{Franciolini:2023pbf,
    author = "Franciolini, Gabriele and Iovino, Junior., Antonio and Vaskonen, Ville and Veermae, Hardi",
    title = "{Recent Gravitational Wave Observation by Pulsar Timing Arrays and Primordial Black Holes: The Importance of Non-Gaussianities}",
    eprint = "2306.17149",
    archivePrefix = "arXiv",
    primaryClass = "astro-ph.CO",
    doi = "10.1103/PhysRevLett.131.201401",
    journal = "Phys. Rev. Lett.",
    volume = "131",
    number = "20",
    pages = "201401",
    year = "2023"
}

@article{Franciolini:2023wjm,
    author = "Franciolini, Gabriele and Racco, Davide and Rompineve, Fabrizio",
    title = "{Footprints of the QCD Crossover on Cosmological Gravitational Waves at Pulsar Timing Arrays}",
    eprint = "2306.17136",
    archivePrefix = "arXiv",
    primaryClass = "astro-ph.CO",
    reportNumber = "CERN-TH-2023-080",
    doi = "10.1103/PhysRevLett.132.081001",
    journal = "Phys. Rev. Lett.",
    volume = "132",
    number = "8",
    pages = "081001",
    year = "2024",
    note = "[Erratum: Phys.Rev.Lett. 133, 189901 (2024)]"
}

@article{Inomata:2023zup,
    author = "Inomata, Keisuke and Kohri, Kazunori and Terada, Takahiro",
    title = "{Detected stochastic gravitational waves and subsolar-mass primordial black holes}",
    eprint = "2306.17834",
    archivePrefix = "arXiv",
    primaryClass = "astro-ph.CO",
    reportNumber = "KEK-TH-2535, KEK-Cosmo-0317, KEK-QUP-2023-0016, CTPU-PTC-23-28",
    doi = "10.1103/PhysRevD.109.063506",
    journal = "Phys. Rev. D",
    volume = "109",
    number = "6",
    pages = "063506",
    year = "2024"
}

@article{Cai:2023dls,
    author = "Cai, Yi-Fu and He, Xin-Chen and Ma, Xiao-Han and Yan, Sheng-Feng and Yuan, Guan-Wen",
    title = "{Limits on scalar-induced gravitational waves from the stochastic background by pulsar timing array observations}",
    eprint = "2306.17822",
    archivePrefix = "arXiv",
    primaryClass = "gr-qc",
    doi = "10.1016/j.scib.2023.10.027",
    journal = "Sci. Bull.",
    volume = "68",
    pages = "2929--2935",
    year = "2023"
}

@article{Wang:2023ost,
    author = "Wang, Sai and Zhao, Zhi-Chao and Li, Jun-Peng and Zhu, Qing-Hua",
    title = "{Implications of pulsar timing array data for scalar-induced gravitational waves and primordial black holes: Primordial non-Gaussianity fNL considered}",
    eprint = "2307.00572",
    archivePrefix = "arXiv",
    primaryClass = "astro-ph.CO",
    reportNumber = "version-01, https://github.com/Zhi-ChaoZhao/sigw{\_}class",
    doi = "10.1103/PhysRevResearch.6.L012060",
    journal = "Phys. Rev. Res.",
    volume = "6",
    number = "1",
    pages = "L012060",
    year = "2024"
}

@article{Liu:2023ymk,
    author = "Liu, Lang and Chen, Zu-Cheng and Huang, Qing-Guo",
    title = "{Implications for the non-Gaussianity of curvature perturbation from pulsar timing arrays}",
    eprint = "2307.01102",
    archivePrefix = "arXiv",
    primaryClass = "astro-ph.CO",
    doi = "10.1103/PhysRevD.109.L061301",
    journal = "Phys. Rev. D",
    volume = "109",
    number = "6",
    pages = "L061301",
    year = "2024"
}

@article{Unal:2023srk,
    author = "Unal, Caner and Papageorgiou, Alexandros and Obata, Ippei",
    title = "{Axion-gauge dynamics during inflation as the origin of pulsar timing array signals and primordial black holes}",
    eprint = "2307.02322",
    archivePrefix = "arXiv",
    primaryClass = "astro-ph.CO",
    doi = "10.1016/j.physletb.2024.138873",
    journal = "Phys. Lett. B",
    volume = "856",
    pages = "138873",
    year = "2024"
}

@article{Figueroa:2023zhu,
    author = "Figueroa, Daniel G. and Pieroni, Mauro and Ricciardone, Angelo and Simakachorn, Peera",
    title = "{Cosmological Background Interpretation of Pulsar Timing Array Data}",
    eprint = "2307.02399",
    archivePrefix = "arXiv",
    primaryClass = "astro-ph.CO",
    reportNumber = "CERN-TH-2023-132",
    doi = "10.1103/PhysRevLett.132.171002",
    journal = "Phys. Rev. Lett.",
    volume = "132",
    number = "17",
    pages = "171002",
    year = "2024"
}

@article{Zhu:2023faa,
    author = "Wang, Sai and Zhao, Zhi-Chao and Zhu, Qing-Hua",
    title = "{Constraints on scalar-induced gravitational waves up to third order from a joint analysis of BBN, CMB, and PTA data}",
    eprint = "2307.03095",
    archivePrefix = "arXiv",
    primaryClass = "astro-ph.CO",
    doi = "10.1103/PhysRevResearch.6.013207",
    journal = "Phys. Rev. Res.",
    volume = "6",
    number = "1",
    pages = "013207",
    year = "2024"
}

@article{Firouzjahi:2023lzg,
    author = "Firouzjahi, Hassan and Talebian, Alireza",
    title = "{Induced gravitational waves from ultra slow-roll inflation and pulsar timing arrays observations}",
    eprint = "2307.03164",
    archivePrefix = "arXiv",
    primaryClass = "gr-qc",
    doi = "10.1088/1475-7516/2023/10/032",
    journal = "JCAP",
    volume = "10",
    pages = "032",
    year = "2023"
}

@article{Li:2023qua,
    author = "Li, Jun-Peng and Wang, Sai and Zhao, Zhi-Chao and Kohri, Kazunori",
    title = "{Primordial non-Gaussianity f $_{NL}$ and anisotropies in scalar-induced gravitational waves}",
    eprint = "2305.19950",
    archivePrefix = "arXiv",
    primaryClass = "astro-ph.CO",
    reportNumber = "KEK-Cosmo-0315, KEK-TH-2531, KEK-QUP-2023-0012, KEK-QUP-2023-0012,
  https://github.com/Zhi-ChaoZhao/sigw{\_}class",
    doi = "10.1088/1475-7516/2023/10/056",
    journal = "JCAP",
    volume = "10",
    pages = "056",
    year = "2023"
}

@article{You:2023rmn,
    author = "You, Zhi-Qiang and Yi, Zhu and Wu, You",
    title = "{Constraints on primordial curvature power spectrum with pulsar timing arrays}",
    eprint = "2307.04419",
    archivePrefix = "arXiv",
    primaryClass = "gr-qc",
    doi = "10.1088/1475-7516/2023/11/065",
    journal = "JCAP",
    volume = "11",
    pages = "065",
    year = "2023"
}

@article{Balaji:2023ehk,
    author = "Balaji, Shyam and Dom{\`e}nech, Guillem and Franciolini, Gabriele",
    title = "{Scalar-induced gravitational wave interpretation of PTA data: the role of scalar fluctuation propagation speed}",
    eprint = "2307.08552",
    archivePrefix = "arXiv",
    primaryClass = "gr-qc",
    doi = "10.1088/1475-7516/2023/10/041",
    journal = "JCAP",
    volume = "10",
    pages = "041",
    year = "2023"
}

@article{HosseiniMansoori:2023mqh,
    author = "Hosseini Mansoori, Seyed Ali and Felegray, Fereshteh and Talebian, Alireza and Sami, Mohammad",
    title = "{PBHs and GWs from {\ensuremath{\mathbb{T}}}$^{2}$-inflation and NANOGrav 15-year data}",
    eprint = "2307.06757",
    archivePrefix = "arXiv",
    primaryClass = "astro-ph.CO",
    doi = "10.1088/1475-7516/2023/08/067",
    journal = "JCAP",
    volume = "08",
    pages = "067",
    year = "2023"
}

@article{Zhao:2023joc,
    author = "Zhu, Qing-Hua and Zhao, Zhi-Chao and Wang, Sai and Zhang, Xin",
    title = "{Unraveling the early universe{\textquoteright}s equation of state and primordial black hole production with PTA, BBN, and CMB observations*}",
    eprint = "2307.13574",
    archivePrefix = "arXiv",
    primaryClass = "astro-ph.CO",
    doi = "10.1088/1674-1137/ad79d5",
    journal = "Chin. Phys. C",
    volume = "48",
    number = "12",
    pages = "125105",
    year = "2024"
}

@article{Liu:2023pau,
    author = "Liu, Lang and Chen, Zu-Cheng and Huang, Qing-Guo",
    title = "{Probing the equation of state of the early Universe with pulsar timing arrays}",
    eprint = "2307.14911",
    archivePrefix = "arXiv",
    primaryClass = "astro-ph.CO",
    doi = "10.1088/1475-7516/2023/11/071",
    journal = "JCAP",
    volume = "11",
    pages = "071",
    year = "2023"
}

@article{Yi:2023tdk,
    author = "Yi, Zhu and You, Zhi-Qiang and Wu, You",
    title = "{Model-independent reconstruction of the primordial curvature power spectrum from PTA data}",
    eprint = "2308.05632",
    archivePrefix = "arXiv",
    primaryClass = "astro-ph.CO",
    doi = "10.1088/1475-7516/2024/01/066",
    journal = "JCAP",
    volume = "01",
    pages = "066",
    year = "2024"
}

@article{Bhaumik:2023wmw,
    author = "Bhaumik, Nilanjandev and Jain, Rajeev Kumar and Lewicki, Marek",
    title = "{Ultralow mass primordial black holes in the early Universe can explain the pulsar timing array signal}",
    eprint = "2308.07912",
    archivePrefix = "arXiv",
    primaryClass = "astro-ph.CO",
    doi = "10.1103/PhysRevD.108.123532",
    journal = "Phys. Rev. D",
    volume = "108",
    number = "12",
    pages = "123532",
    year = "2023"
}

@article{Choudhury:2023hfm,
    author = "Choudhury, Sayantan and Karde, Ahaskar and Panda, Sudhakar and Sami, M.",
    title = "{Scalar induced gravity waves from ultra slow-roll galileon inflation}",
    eprint = "2308.09273",
    archivePrefix = "arXiv",
    primaryClass = "astro-ph.CO",
    doi = "10.1016/j.nuclphysb.2024.116678",
    journal = "Nucl. Phys. B",
    volume = "1007",
    pages = "116678",
    year = "2024"
}

@article{Yi:2023npi,
    author = "Yi, Zhu and You, Zhi-Qiang and Wu, You and Chen, Zu-Cheng and Liu, Lang",
    title = "{Exploring the NANOGrav signal and planet-mass primordial black holes through Higgs inflation}",
    eprint = "2308.14688",
    archivePrefix = "arXiv",
    primaryClass = "astro-ph.CO",
    doi = "10.1088/1475-7516/2024/06/043",
    journal = "JCAP",
    volume = "06",
    pages = "043",
    year = "2024"
}

@article{Harigaya:2023pmw,
    author = "Harigaya, Keisuke and Inomata, Keisuke and Terada, Takahiro",
    title = "{Induced gravitational waves with kination era for recent pulsar timing array signals}",
    eprint = "2309.00228",
    archivePrefix = "arXiv",
    primaryClass = "astro-ph.CO",
    reportNumber = "CTPU-PTC-23-40",
    doi = "10.1103/PhysRevD.108.123538",
    journal = "Phys. Rev. D",
    volume = "108",
    number = "12",
    pages = "123538",
    year = "2023"
}

@article{Jin:2023wri,
    author = "Jin, Jia-Heng and Chen, Zu-Cheng and Yi, Zhu and You, Zhi-Qiang and Liu, Lang and Wu, You",
    title = "{Confronting sound speed resonance with pulsar timing arrays}",
    eprint = "2307.08687",
    archivePrefix = "arXiv",
    primaryClass = "astro-ph.CO",
    doi = "10.1088/1475-7516/2023/09/016",
    journal = "JCAP",
    volume = "09",
    pages = "016",
    year = "2023"
}

@article{Cannizzaro:2023mgc,
    author = "Cannizzaro, Enrico and Franciolini, Gabriele and Pani, Paolo",
    title = "{Novel tests of gravity using nano-Hertz stochastic gravitational-wave background signals}",
    eprint = "2307.11665",
    archivePrefix = "arXiv",
    primaryClass = "gr-qc",
    doi = "10.1088/1475-7516/2024/04/056",
    journal = "JCAP",
    volume = "04",
    pages = "056",
    year = "2024"
}

@article{Zhang:2023nrs,
    author = "Zhang, Zhao and Cai, Chengfeng and Su, Yu-Hang and Wang, Shiyu and Yu, Zhao-Huan and Zhang, Hong-Hao",
    title = "{Nano-Hertz gravitational waves from collapsing domain walls associated with freeze-in dark matter in light of pulsar timing array observations}",
    eprint = "2307.11495",
    archivePrefix = "arXiv",
    primaryClass = "hep-ph",
    doi = "10.1103/PhysRevD.108.095037",
    journal = "Phys. Rev. D",
    volume = "108",
    number = "9",
    pages = "095037",
    year = "2023"
}

@article{Liu:2023hpw,
    author = "Liu, Lang and Wu, You and Chen, Zu-Cheng",
    title = "{Simultaneously probing the sound speed and equation of state of the early Universe with pulsar timing arrays}",
    eprint = "2310.16500",
    archivePrefix = "arXiv",
    primaryClass = "astro-ph.CO",
    doi = "10.1088/1475-7516/2024/04/011",
    journal = "JCAP",
    volume = "04",
    pages = "011",
    year = "2024"
}

@article{Choudhury:2023fwk,
    author = "Choudhury, Sayantan and Dey, Kritartha and Karde, Ahaskar and Panda, Sudhakar and Sami, M.",
    title = "{Primordial non-Gaussianity as a saviour for PBH overproduction in SIGWs generated by pulsar timing arrays for Galileon inflation}",
    eprint = "2310.11034",
    archivePrefix = "arXiv",
    primaryClass = "astro-ph.CO",
    doi = "10.1016/j.physletb.2024.138925",
    journal = "Phys. Lett. B",
    volume = "856",
    pages = "138925",
    year = "2024"
}

@article{Tagliazucchi:2023dai,
    author = "Tagliazucchi, Matteo and Braglia, Matteo and Finelli, Fabio and Pieroni, Mauro",
    title = "{Quest for CMB spectral distortions to probe the scalar-induced gravitational wave background interpretation of pulsar timing array data}",
    eprint = "2310.08527",
    archivePrefix = "arXiv",
    primaryClass = "astro-ph.CO",
    reportNumber = "CERN-TH-2023-191",
    doi = "10.1103/PhysRevD.111.L021305",
    journal = "Phys. Rev. D",
    volume = "111",
    number = "2",
    pages = "L021305",
    year = "2025"
}

@article{Basilakos:2023jvp,
    author = "Basilakos, Spyros and Nanopoulos, Dimitri V. and Papanikolaou, Theodoros and Saridakis, Emmanuel N. and Tzerefos, Charalampos",
    title = "{Induced gravitational waves from flipped SU(5) superstring theory at nHz}",
    eprint = "2309.15820",
    archivePrefix = "arXiv",
    primaryClass = "astro-ph.CO",
    doi = "10.1016/j.physletb.2024.138446",
    journal = "Phys. Lett. B",
    volume = "849",
    pages = "138446",
    year = "2024"
}

@article{Inomata:2023drn,
    author = "Inomata, Keisuke and Kawasaki, Masahiro and Mukaida, Kyohei and Yanagida, Tsutomu T.",
    title = "{Axion curvaton model for the gravitational waves observed by pulsar timing arrays}",
    eprint = "2309.11398",
    archivePrefix = "arXiv",
    primaryClass = "astro-ph.CO",
    reportNumber = "KEK-TH-2554, KEK-Cosmo-0324",
    doi = "10.1103/PhysRevD.109.043508",
    journal = "Phys. Rev. D",
    volume = "109",
    number = "4",
    pages = "043508",
    year = "2024"
}

@article{Li:2023xtl,
    author = "Li, Jun-Peng and Wang, Sai and Zhao, Zhi-Chao and Kohri, Kazunori",
    title = "{Complete analysis of the background and anisotropies of scalar-induced gravitational waves: primordial non-Gaussianity f $_{NL}$ and g $_{NL}$ considered}",
    eprint = "2309.07792",
    archivePrefix = "arXiv",
    primaryClass = "astro-ph.CO",
    reportNumber = "KEK-Cosmo-0326, KEK-TH-2556, KEK-QUP-2023-0024, KEK-QUP-2023-0024,
  https://github.com/Zhi-ChaoZhao/sigw{\_}class",
    doi = "10.1088/1475-7516/2024/06/039",
    journal = "JCAP",
    volume = "06",
    pages = "039",
    year = "2024"
}

@article{Domenech:2023dxx,
    author = "Dom{\`e}nech, Guillem and Vargas, Gerson and Vargas, Te{\'o}filo",
    title = "{An exact model for enhancing/suppressing primordial fluctuations}",
    eprint = "2309.05750",
    archivePrefix = "arXiv",
    primaryClass = "astro-ph.CO",
    doi = "10.1088/1475-7516/2024/03/002",
    journal = "JCAP",
    volume = "03",
    pages = "002",
    year = "2024"
}

@article{Gangopadhyay:2023qjr,
    author = "Gangopadhyay, M. R. and Godithi, V. V. and Inui, R. and Ichiki, K. and Kajino, T. and Manusankar, A. and Mathews, G. J. and Yogesh",
    title = "{Is the NANOGrav detection evidence of resonant particle creation during inflation?}",
    eprint = "2309.03101",
    archivePrefix = "arXiv",
    primaryClass = "astro-ph.CO",
    doi = "10.1016/j.jheap.2025.100358",
    journal = "JHEAp",
    volume = "47",
    pages = "100358",
    year = "2025"
}

@article{Cyr:2023pgw,
    author = "Cyr, Bryce and Kite, Thomas and Chluba, Jens and Hill, J. Colin and Jeong, Donghui and Acharya, Sandeep Kumar and Bolliet, Boris and Patil, Subodh P.",
    title = "{Disentangling the primordial nature of stochastic gravitational wave backgrounds with CMB spectral distortions}",
    eprint = "2309.02366",
    archivePrefix = "arXiv",
    primaryClass = "astro-ph.CO",
    doi = "10.1093/mnras/stad3861",
    journal = "Mon. Not. Roy. Astron. Soc.",
    volume = "528",
    number = "1",
    pages = "883--897",
    year = "2024"
}

@article{Chen:2024fir,
    author = "Chen, Zu-Cheng and Li, Jun and Liu, Lang and Yi, Zhu",
    title = "{Probing the speed of scalar-induced gravitational waves with pulsar timing arrays}",
    eprint = "2401.09818",
    archivePrefix = "arXiv",
    primaryClass = "gr-qc",
    doi = "10.1103/PhysRevD.109.L101302",
    journal = "Phys. Rev. D",
    volume = "109",
    number = "10",
    pages = "L101302",
    year = "2024"
}

@article{Chen:2024twp,
    author = "Chen, Zu-Cheng and Liu, Lang",
    title = "{Can we distinguish between adiabatic and isocurvature fluctuations with pulsar timing arrays?}",
    eprint = "2402.16781",
    archivePrefix = "arXiv",
    primaryClass = "astro-ph.CO",
    doi = "10.1007/s11433-025-2614-0",
    journal = "Sci. China Phys. Mech. Astron.",
    volume = "68",
    number = "5",
    pages = "250412",
    year = "2025"
}

@article{Choudhury:2023fjs,
    author = "Choudhury, Sayantan and Dey, Kritartha and Karde, Ahaskar",
    title = "{Untangling PBH Overproduction in w$w$-SIGWs Generated by Pulsar Timing Arrays for MST-EFT of Single Field Inflation}",
    eprint = "2311.15065",
    archivePrefix = "arXiv",
    primaryClass = "astro-ph.CO",
    doi = "10.1002/prop.70067",
    journal = "Fortsch. Phys.",
    volume = "74",
    number = "1",
    pages = "e70067",
    year = "2026"
}

@article{Choudhury:2024one,
    author = "Choudhury, Sayantan and Karde, Ahaskar and Panda, Sudhakar and Sami, M.",
    title = "{Realisation of the ultra-slow roll phase in Galileon inflation and PBH overproduction}",
    eprint = "2401.10925",
    archivePrefix = "arXiv",
    primaryClass = "astro-ph.CO",
    doi = "10.1088/1475-7516/2024/07/034",
    journal = "JCAP",
    volume = "07",
    pages = "034",
    year = "2024"
}

@article{Cai:2025ksu,
    author = "Cai, Yi-Fu and Du, Peizhi and Zhong, Jiahang",
    title = "{Isocurvature induced gravitational waves at Pulsar Timing Arrays}",
    eprint = "2512.08301",
    archivePrefix = "arXiv",
    primaryClass = "astro-ph.CO",
    doi = "10.1088/1475-7516/2026/05/066",
    journal = "JCAP",
    volume = "05",
    pages = "066",
    year = "2026"
}

@article{Pi:2020otn,
    author = "Pi, Shi and Sasaki, Misao",
    title = "{Gravitational Waves Induced by Scalar Perturbations with a Lognormal Peak}",
    eprint = "2005.12306",
    archivePrefix = "arXiv",
    primaryClass = "gr-qc",
    reportNumber = "YITP-20-75, YITP-75, IPMU20-0054",
    doi = "10.1088/1475-7516/2020/09/037",
    journal = "JCAP",
    volume = "09",
    pages = "037",
    year = "2020"
}

@article{Li:2024lxx,
    author = "Li, Chong-Zhi and Yuan, Chen and Huang, Qing-guo",
    title = "{Gravitational waves induced by scalar perturbations with a broken power-law peak}",
    eprint = "2407.12914",
    archivePrefix = "arXiv",
    primaryClass = "gr-qc",
    doi = "10.1088/1475-7516/2025/01/067",
    journal = "JCAP",
    volume = "01",
    pages = "067",
    year = "2025"
}

@article{Cai:2019cdl,
    author = "Cai, Rong-Gen and Pi, Shi and Sasaki, Misao",
    title = "{Universal infrared scaling of gravitational wave background spectra}",
    eprint = "1909.13728",
    archivePrefix = "arXiv",
    primaryClass = "astro-ph.CO",
    reportNumber = "IPMU19-0135, YITP-19-88",
    doi = "10.1103/PhysRevD.102.083528",
    journal = "Phys. Rev. D",
    volume = "102",
    number = "8",
    pages = "083528",
    year = "2020"
}

@article{Yuan:2019wwo,
    author = "Yuan, Chen and Chen, Zu-Cheng and Huang, Qing-Guo",
    title = "{Log-dependent slope of scalar induced gravitational waves in the infrared regions}",
    eprint = "1910.09099",
    archivePrefix = "arXiv",
    primaryClass = "astro-ph.CO",
    doi = "10.1103/PhysRevD.101.043019",
    journal = "Phys. Rev. D",
    volume = "101",
    number = "4",
    pages = "4",
    year = "2020"
}

@article{Inomata:2019yww,
    author = "Inomata, Keisuke and Terada, Takahiro",
    title = "{Gauge Independence of Induced Gravitational Waves}",
    eprint = "1912.00785",
    archivePrefix = "arXiv",
    primaryClass = "gr-qc",
    reportNumber = "IPMU 19-0178, CTPU-PTC-19-36",
    doi = "10.1103/PhysRevD.101.023523",
    journal = "Phys. Rev. D",
    volume = "101",
    number = "2",
    pages = "023523",
    year = "2020"
}

@article{DeLuca:2019ufz,
    author = "De Luca, V. and Franciolini, G. and Kehagias, A. and Riotto, A.",
    title = "{On the Gauge Invariance of Cosmological Gravitational Waves}",
    eprint = "1911.09689",
    archivePrefix = "arXiv",
    primaryClass = "gr-qc",
    doi = "10.1088/1475-7516/2020/03/014",
    journal = "JCAP",
    volume = "03",
    pages = "014",
    year = "2020"
}

@article{Yuan:2019fwv,
    author = "Yuan, Chen and Chen, Zu-Cheng and Huang, Qing-Guo",
    title = "{Scalar Induced Gravitational Waves in Different Gauges}",
    eprint = "1912.00885",
    archivePrefix = "arXiv",
    primaryClass = "astro-ph.CO",
    doi = "10.1103/PhysRevD.101.063018",
    journal = "Phys. Rev. D",
    volume = "101",
    number = "6",
    pages = "6",
    year = "2020"
}

@article{Hwang:2017oxa,
    author = "Hwang, Jai-Chan and Jeong, Donghui and Noh, Hyerim",
    title = "{Gauge dependence of gravitational waves generated from scalar perturbations}",
    eprint = "1704.03500",
    archivePrefix = "arXiv",
    primaryClass = "astro-ph.CO",
    doi = "10.3847/1538-4357/aa74be",
    journal = "Astrophys. J.",
    volume = "842",
    number = "1",
    pages = "46",
    year = "2017"
}

@article{Tomikawa:2019tvi,
    author = "Tomikawa, Keitaro and Kobayashi, Tsutomu",
    title = "{Gauge dependence of gravitational waves generated at second order from scalar perturbations}",
    eprint = "1910.01880",
    archivePrefix = "arXiv",
    primaryClass = "gr-qc",
    reportNumber = "RUP-19-27",
    doi = "10.1103/PhysRevD.101.083529",
    journal = "Phys. Rev. D",
    volume = "101",
    number = "8",
    pages = "083529",
    year = "2020"
}

@article{Domenech:2020xin,
    author = "Dom{\`e}nech, Guillem and Sasaki, Misao",
    title = "{Approximate gauge independence of the induced gravitational wave spectrum}",
    eprint = "2012.14016",
    archivePrefix = "arXiv",
    primaryClass = "gr-qc",
    reportNumber = "YITP-20-163",
    doi = "10.1103/PhysRevD.103.063531",
    journal = "Phys. Rev. D",
    volume = "103",
    number = "6",
    pages = "063531",
    year = "2021"
}

@article{Yuan:2024qfz,
    author = "Yuan, Chen and Chen, Zu-Cheng and Liu, Lang",
    title = "{Gauge dependence of gravitational waves induced by primordial isocurvature fluctuations}",
    eprint = "2410.18996",
    archivePrefix = "arXiv",
    primaryClass = "gr-qc",
    doi = "10.1103/PhysRevD.111.103528",
    journal = "Phys. Rev. D",
    volume = "111",
    number = "10",
    pages = "103528",
    year = "2025"
}

@article{Yuan:2025seu,
    author = "Yuan, Chen and Lu, Yizhou and Chen, Zu-Cheng and Liu, Lang",
    title = "{On the gauge invariance of secondary gravitational~waves}",
    eprint = "2501.13691",
    archivePrefix = "arXiv",
    primaryClass = "astro-ph.CO",
    doi = "10.1088/1475-7516/2025/07/016",
    journal = "JCAP",
    volume = "07",
    pages = "016",
    year = "2025"
}

@article{Domenech:2025ccu,
    author = "Dom{\`e}nech, Guillem and Pi, Shi and Wang, Ao",
    title = "{Observable Gravitational Wave Strain at Second Order}",
    eprint = "2512.15704",
    archivePrefix = "arXiv",
    primaryClass = "gr-qc",
    doi = "10.1103/pwbs-xwrh",
    journal = "Phys. Rev. Lett.",
    volume = "136",
    number = "22",
    pages = "221402",
    year = "2026"
}

@article{Unal:2018yaa,
    author = "Unal, Caner",
    title = "{Imprints of Primordial Non-Gaussianity on Gravitational Wave Spectrum}",
    eprint = "1811.09151",
    archivePrefix = "arXiv",
    primaryClass = "astro-ph.CO",
    doi = "10.1103/PhysRevD.99.041301",
    journal = "Phys. Rev. D",
    volume = "99",
    number = "4",
    pages = "041301",
    year = "2019"
}

@article{Atal:2021jyo,
    author = "Atal, Vicente and Dom{\`e}nech, Guillem",
    title = "{Probing non-Gaussianities with the high frequency tail of induced gravitational waves}",
    eprint = "2103.01056",
    archivePrefix = "arXiv",
    primaryClass = "astro-ph.CO",
    doi = "10.1088/1475-7516/2021/06/001",
    journal = "JCAP",
    volume = "06",
    pages = "001",
    year = "2021",
    note = "[Erratum: JCAP 10, E01 (2023)]"
}

@article{Franciolini:2018vbk,
    author = "Franciolini, G. and Kehagias, A. and Matarrese, S. and Riotto, A.",
    title = "{Primordial Black Holes from Inflation and non-Gaussianity}",
    eprint = "1801.09415",
    archivePrefix = "arXiv",
    primaryClass = "astro-ph.CO",
    doi = "10.1088/1475-7516/2018/03/016",
    journal = "JCAP",
    volume = "03",
    pages = "016",
    year = "2018"
}

@article{Atal:2019cdz,
    author = "Atal, Vicente and Garriga, Jaume and Marcos-Caballero, Airam",
    title = "{Primordial black hole formation with non-Gaussian curvature perturbations}",
    eprint = "1905.13202",
    archivePrefix = "arXiv",
    primaryClass = "astro-ph.CO",
    doi = "10.1088/1475-7516/2019/09/073",
    journal = "JCAP",
    volume = "09",
    pages = "073",
    year = "2019"
}

@article{Perna:2026szd,
    author = "Perna, Gabriele and Dom{\`e}nech, Guillem",
    title = "{Probing non-Gaussianity during reheating with SIGW in the LISA band}",
    eprint = "2604.08493",
    archivePrefix = "arXiv",
    primaryClass = "astro-ph.CO",
    month = "4",
    year = "2026"
}

\end{document}